\documentclass[times,twocolumn,final,authoryear]{elsarticle}

\usepackage{jasr}
\usepackage{framed,multirow}

\usepackage{amssymb}
\usepackage{latexsym}

\usepackage{amsmath}
\usepackage{color,colortbl}

\usepackage[switch]{lineno}

\usepackage{url}
\usepackage{xcolor}
\definecolor{newcolor}{rgb}{.8,.349,.1}

\usepackage[citebordercolor=white]{hyperref}

\usepackage[normalem]{ulem}

\journal{Advances in Space Research}

\begin{document}

\verso{E.A.Godenko \textit{etal}}

\begin{frontmatter}

\title{Dynamical charging of interstellar dust particles in the heliosphere}

\author[1,2,3]{E.A.Godenko \corref{cor1}}
\cortext[cor1]{Corresponding author: 
  Tel.: +7-926-679-0812;}
\ead{godenko.egor@yandex.ru}
\author[1,2,3]{V.V.Izmodenov}

\address[1]{Ishlinsky Institute for Problems in Mechanics, Russian Academy of Sciences, pr. Vernadskogo 101-1, Moscow, 119526 Russia}
\address[2]{Lomonosov Moscow State University, Moscow Center for Fundamental and Applied Mathematics, GSP-1, Leninskie Gory, Moscow, 119991 Russia}
\address[3]{Space Research Institute, Russian Academy of Sciences, Profsoyuznaya Str. 84/32, Moscow, 117335 Russia}

\received{19 April 2023}
\finalform{???}
\accepted{??}
\availableonline{???}
\communicated{???}

\begin{abstract}
Interstellar dust (ISD) particles penetrate the solar system due to the relative motion of the Sun and the local interstellar cloud. Before entering the heliosphere, they pass through the heliospheric interface -- the region of the solar wind interaction with the interstellar plasma. The size distribution and number density of dust grains are modified in the interface essentially. The modification depends on the charging of the dust particles along their trajectories. In this paper, we present modeling results of the charging of ISD particles passing through the heliospheric interface. The main physical processes responsible for the charging within the heliospheric conditions are the sticking of primary plasma particles, secondary electron emission, photoemission, and the effects of cosmic ray electrons. We consider two methods to calculate the electric charge of ISD particles based on (1) the classic steady-state assumption that the charge depends only on local plasma and radiation conditions and (2) the dynamical computation of charge along the particle trajectory. We demonstrate that the steady-state assumption is quite justified to model trajectories and number density distributions of relatively big ISD grains (radius of 100 nm and larger) penetrating the heliosphere. The estimates show that ISD grains of these sizes require less than $0.25$ years (distance of $\approx 1$ au) after transition from the LISM into the heliosphere to reach an equilibrium. For small particles (radius of 10 nm), the dynamical computation of charge influences the trajectories and modifies the number density substantially. The dust density accumulations are distributed within a more elongated region along the heliopause in case of dynamically changed charge as compared with the use of a steady-state charge approximation. As a result, the magnitudes of number density at the points of density features differ several times between the results obtained by the two considered approaches.
\end{abstract}


\end{frontmatter}


\section{Introduction} \label{sec: introduction}

The Sun moves through the local interstellar medium (LISM) with a bulk velocity of $\sim 26.4$ km s$^{-1}$ \citep{witte_2004}. As a result of interaction between the solar wind and the incoming flow of partially ionized LISM plasma, a complex structure consisting of two shock waves -- the heliospheric termination shock and the bow shock in LISM -- and tangential discontinuity -- the heliopause -- forms. This structure is called the heliospheric interface. For the first time, such a structure was proposed by \cite{baranov_1970}. Crossings of the heliospheric termination shock and the heliopause by Voyager 1 and 2 confirmed this theory, while the existence of bow shock is under the question so far \citep[see, e.g.,][]{izmodenov_2009, mccomas_2012, zank_2013}. Now several scientific groups have been developing global models of the heliosphere \citep[see, e.g.,][]{izmodenov_alexashov_2020, pogorelov_2021, opher_2020} with the use of different approaches and numerical methods.

In addition to the charged and neutral components, the LISM plasma also contains dust particles, as was undoubtedly detected for the first time by the Ulysses measurements \citep{grun_1993}. Interstellar dust (ISD) particles are solid grains whose sizes are approximately from a few nanometers up to tens of micrometers \citep[see, e.g.,][]{mann_2010, sterken_2022}. The size distribution of ISD grains in the interstellar medium (ISM) was obtained using the astronomical observations on the extinction and polarization of light by dust particles \citep[see, e.g.,][]{mathis_1977, weingartner_2001, wang_2015}. The ISD grains can penetrate deep inside the heliosphere due to the relative motion of the Sun in the LISM. It opens a possibility to study them by in situ measurements. \cite{kruger_2015}, \cite{strub_2015}, \cite{sterken_2015} carried out a detailed analysis of the Ulysses dataset, but some questions on these data are still open. The whole dataset from Ulysses could not be understood in the frame of a single simulation run: some features of this dataset could be reproduced by relatively big grains ($a > 0.2$ $\mu$m, the data before 2003), and other ones are by small grains ($a < 0.2$ $\mu $m, the data after 2005). The exploration of the chemical composition of these grains has been made possible by the time-of-flight mass spectrometer mounted on the Cassini spacecraft \citep{altobelli_2016} and by the Stardust sample return mission \citep{westphal_2014}. Mostly rock-forming elements (Mg, Si, Fe, Ca) were observed among the constituents of the detected ISD particles. Carbon-rich grains were not registered, although the astronomical observations indicate their existence in the ISM. \cite{sterken_2019} presented a detailed review of the ISD measurements.

The interaction between the solar wind and LISM plasma creates a specific environment that deflects the straight-line trajectories of ISD grains. However, since the magnitudes of solar gravitational and radiation pressure forces are inversely proportional to the squared heliodistance, they become strong only near the Sun. Thus, the electromagnetic forces govern the dynamics of ISD particles in the heliospheric interface. One should note that for the stars with extremely intensive radiation fields, the radiation pressure force persists dominant far from the star \citep[see][]{katushkina_2019}, but this is not the case for the Sun. The magnitude of the electromagnetic force depends on the electric charge of a dust grain, which is affected by many physical processes occurring in the global heliosphere:  (1) sticking of particles from surrounding plasma on the surface of a dust grain, (2) emission of secondary electrons from the grain due to the bombardment by high-energy primary plasma particles, (3) different types of photoelectric emission: primary, Auger, secondary, photodetachment \citep{weingartner_2006}, and (4) the effects of cosmic rays \citep{ivlev_2015}. Currents corresponding to these processes are determined by the conditions of surrounding plasma, radiation, and geometrical (size, shape) and chemical properties of examined dust grains. Other physical processes, for example, thermionic emission and field emission, can be neglected in the regions of interest \citep{kimura_mann_1999}.

The theoretical study of ISD charging began before their unambiguous detection in the measurements of the Ulysses spacecraft. \cite{draine_salpeter_1979} presented one of the first works on dust charging within the hot plasma environment. They reexamined the sticking of primary plasma particles, secondary electron emission, and photoemission and performed computations of the steady-state charge for certain ISM conditions. They also provided expressions for secondary electron emission yield accounting for the spherical curvature of a dust grain surface, which is especially important for small dust grains. \cite{draine_sutin_1987} considered a statistical kinetic approach to study the equilibrium charging of dust grains based on the analysis of charge probability distribution function instead of the calculation of steady-state charge. They considered only collisional charging by the particles from surrounding plasma. \cite{horanyi_1996} provided the summary of knowledge on dust charging at that moment. In the presented charging model, they took account of the relative velocity between a dust grain and bulk plasma when computing the rates of impinging plasma particles that could be important in the regions where dust grains coming from the ISM meet the supersonic solar wind. Besides, a more detailed approach to computing the rates of secondary electrons was considered by averaging secondary electron yield over the Maxwellian thermal distribution \citep{meyer-vernet_1982}. They applied the resulting charging model to different conditions in the solar system. \cite{kimura_mann_1998} applied an assumption of steady-state surface charge potential and performed computations for two species of ISD grains (carbon and silicate) at different positions in the heliosphere using as input the distributions of plasma parameters from \cite{pauls_zank_1996}. They showed that the steady-state potential strongly depends on the heliocentric distance and, in particular, increases in the inner heliospheric interface due to increasing rates of secondary electron emission. They also investigated how dust grain mass and temporal variations of solar wind parameters influence the value of steady-state potential. In subsequent works, they provided updated estimates for secondary electron emission \citep{kimura_mann_1999} and photoelectric \citep{kimura_2016} yields. \cite{slavin_2012} used a charging model from \cite{weingartner_2006} and the heliospheric model from \cite{pogorelov_2008} to study the trajectories of ISD grains. Figure 2 of their paper shows a small-particle effect (i.e., the increase of steady-state potential with decreasing size) caused by the dependence of secondary electron emission and photoelectric yields on the grain size. They also demonstrated that the heliospheric boundaries filter out small ISD grains from the penetration inside the heliosphere. However, they did not perform a detailed analysis of ISD charging in the heliosphere because they focused mostly on the trajectories and distributions of ISD grains.  \cite{alexashov_2016} also performed the computations of ISD steady-state potential using the distributions of plasma parameters from \cite{izmodenov_alexashov_2015} and applied them for the studying of dust density distributions. Since they used formulae for photoelectric emission and secondary electron emission currents \citep{goertz_1989}, which do not take account of the curvature of a dust grain surface, the resulting potential does not depend on the size of dust grains in their charging model. \cite{ibanez_2019} studied dust charging in the ISM for some baseline cases of surrounding conditions. They considered the effects related to cosmic ray electrons as the collection of these electrons by a dust grain and extra radiation induced by these electrons within dense molecular clouds \citep{ivlev_2015}. They calculated the charge probability distribution of dust grains using the approach from \cite{draine_sutin_1987} and demonstrated how this distribution is affected by cosmic ray electrons.

In our previous work \citep{godenko_2021b}, we applied the relatively advanced charging model to compute the density distributions of ISD grains of different sizes. However, we did not go deep in the nature of dust charging. Here, we present a detailed study of ISD charging in the global heliosphere. For computations, we apply the distributions of plasma parameters obtained from the global 3D kinetic-MHD heliospheric model of \cite{izmodenov_alexashov_2020}. At first, we show what processes govern the charging in different regions of the heliospheric interface and compute the steady-state surface charge potential. However, \cite{kimura_mann_1998} proposed that the steady-state assumption is not valid for the smallest interstellar grains. That is why we perform the computations of the dynamically changed charge using the distinct differential equation instead of steady-state charge approximation. We demonstrate for the first time how the ISD charge evolves across the heliopause for particles of different sizes and study the effects on density distribution for small ISD grains.

The structure of the paper is as follows. In Section \ref{section: model}, we describe the approaches used for the ISD charge modeling. In Section \ref{section: currents}, we discuss the physical processes taken into account in the modeling and present the expressions applied to compute the corresponding currents. In Section \ref{section: results}, we demonstrate the modeling results: the currents as functions of surface charge potential and the values of steady-state potential. In Section \ref{section: dynamical_effects}, we investigate the influence of dynamically changed charge on ISD trajectories and density distributions. Section \ref{section: conclusions} concludes the paper.

\section{Charging Models} \label{section: model}

The evolution of dust grain charge, $Q$, can be expressed, generally, in the following form:

\begin{equation} \label{formula: general_equation}
    \frac{dQ}{dt} = \sum\limits_{k}J_k,
\end{equation}

\noindent
where $J_k$ is the current corresponding to the $k$-th physical process of dust grain charging. In this paper, we consider the following processes:

\begin{itemize}
    \item sticking of thermal plasma protons and electrons onto dust grains (Section \ref{subsection: sticking});
    \item the secondary electron emission induced due to the bombardments by high-energy impinging plasma particles (Section \ref{subsection: secondary_electron_emission});
    \item photoelectric emission (Section \ref{subsection: photoelectric emission});
    \item sticking of the cosmic ray electrons and secondary electron emission induced by them (Section \ref{subsection: cr_effects}).
\end{itemize}

To model the trajectories of ISD grains, one should solve equations of their motion together with equation (\ref{formula: general_equation}). However, this approach is ineffective, from the numerical point of view, because characteristic time-scales of charging in equation (\ref{formula: general_equation}) are much smaller than the time-scales of the trajectory equations. Therefore, in the numerical solution, the integration time-step required for equation (\ref{formula: general_equation}) is much smaller than the time-step required by trajectory equations. It makes direct numerical integration computationally intensive and ineffective.

To resolve this issue, one can employ an assumption of steady-state charge. The steady-state charge is the charge for which the right-hand side of the equation (\ref{formula: general_equation}) equals zero. This assumption comes from the fact that the characteristic time of charge evolution is less than the time scales of ISD motion in the heliosphere by one or several orders of magnitude. It enables us to increase the efficiency of the ISD trajectory modeling using the preliminary computations of steady-state charge at any point of interest. Thus, one can compute the values of steady-state charge from the algebraic equation:

\begin{equation} \label{formula: equilibrium_charge_equation}
    \sum\limits_{k}J_k(Q) = 0,
\end{equation}

\noindent
relative to the variable $Q$ (or $U$ if we introduce currents $J_k$ in terms of surface charge potential). Below, in Section \ref{section: dynamical_effects}, we show how accurate dynamical solving of the differential equation for charge modifies the trajectories and number density distributions of ISD grains compared to the case of steady-state assumption.

\section{Currents} \label{section: currents}

In this Section, we briefly discuss the physical aspects of mentioned processes and present formulae used in this study to calculate the corresponding currents. Hereafter, we assume that dust particles have a spherical shape of radius $a$, although they are probably non-spherical actually \citep[for the charging of irregularly shaped dust aggregates, see, e.g.,][]{ma_2013}. While there are some uncertainties in the chemical composition of ISD grains, throughout this paper we consider astronomical silicates MgFeSiO$_4$ with a bulk density of $\rho = 3.5$ g cm$^{-3}$ intermediate between the values for crystalline forsterite and fayalite as was adopted in \cite{weingartner_2006}. However, one can use our charging model for any other material by just varying the values of some physical quantities , and, for comparison, we also demonstrate the estimates of steady-state charge (Section \ref{subsection: results_steady_state}) for carbonaceous (C) grains with a bulk density of $2.24$ g cm$^{-3}$ \citep{weingartner_2006}. For spherical grains, the mass is $m = \frac{4}{3}\pi\rho a^3$, and the charge is $Q = 4\pi \varepsilon_0 U a$, where $U$ is the surface charge potential (assuming that surrounding reference potential equals 0 V), $\varepsilon_0$ is the vacuum permittivity. Thus, we present the modeling results primarily in terms of surface charge potential $U$ instead of charge $Q$ for convenience because the orders of magnitude of $U$ are the same for dust particles of different sizes (as opposed to $Q$).

\subsection{Sticking of plasma particles} \label{subsection: sticking}

Dust grains moving through a hot plasma undergo collisions with plasma particles. In the model of the heliosphere used in this paper \citep[for details, see Appendix A in ][]{izmodenov_alexashov_2020}, plasma is assumed locally Maxwellian and consists of protons and electrons. Pick-up protons are not considered as an individual component of plasma in this model, but they contribute to thermal protons. While reaching the dust grain surface, plasma particles can stick to it. The estimates on the probability of sticking $s_i$, or sticking coefficient, are obtained from the experimental results \citep{weingartner_2001}, and we adopt the corresponding empirical expressions (see \ref{appendix: sticking_coefficient}). One can express the current of plasma particles of sort $i$ ($i$ runs through $p, el$ -- proton and electron, correspondingly) sticking to the surface of a dust grain by the number of collisions per unit of time between plasma particles and a dust grain:

\begin{equation} \label{formula: sticking_general_integral}
    J^{st}_{i} = s_i q_i e
    \int\limits_{0}^{\infty} \int\limits_{0}^{\pi}\int\limits_{0}^{2\pi}\sigma_{i}(v)f_{i}(v, \vartheta)v^3\sin{\vartheta}d{\varphi}d{\vartheta}dv,
\end{equation}

\noindent
where $q_i e$ is the charge of corresponding plasma particles ($q_p = 1, \: q_{el} = -1, \: e$ is the absolute measure of the elementary charge), $\sigma_i(v)$ is the collisional cross section (see \ref{appendix: collisional_cross_section} for details), $f_i(v, \theta)$ is the Maxwellian distribution function:

\begin{equation*}
    f_i(v, \vartheta) = n_i\left(\frac{m_i}{2\pi k T_i}
    \right)^{\frac{3}{2}}\exp\left[-\frac{m_i}{2kT_i}(v^2 + w^2 - 2v w\cos{\vartheta})\right]
\end{equation*}

\noindent
where $n_i, m_i, T_i$ are the number density, particle mass, and temperature of $i$-th plasma component, respectively, $k$ is the Boltzmann constant, $w$ is the plasma bulk velocity relative to a dust grain, and the integration is performed over the velocity components represented in a spherical coordinate system associated with a dust grain. 3D integral in (\ref{formula: sticking_general_integral}) can be reduced to 1D integral after performing the analytical integration over $\vartheta$ and $\varphi$ variables (see \ref{appendix: reduced_collisional_integrals}). Moreover, \cite{kimura_mann_1998} showed that for a simple Coulomb cross section (eq. (5) in their paper), integral (\ref{formula: sticking_general_integral}) could be represented in an almost analytical form with the use of error function (see eq. (7)-(13) in their paper). However, here we use a more advanced expression for a collisional cross section \citep{draine_sutin_1987}, which takes account of the polarization of a dust grain induced by an electric field of impinging plasma particles.

\subsection{Secondary electron emission} \label{subsection: secondary_electron_emission}

\begin{figure}
    \center{\includegraphics[width=1\linewidth]{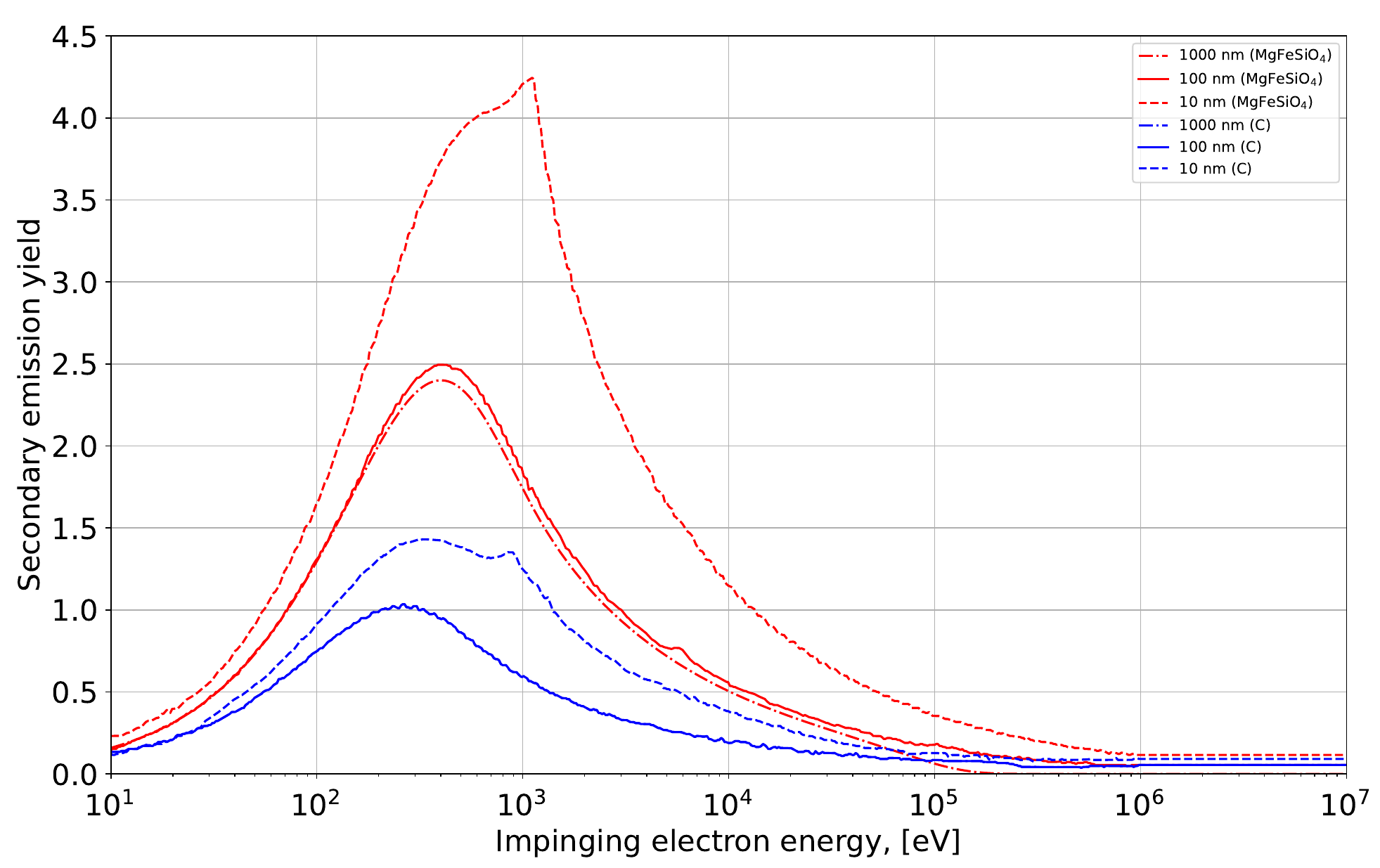}}
    \caption{The secondary electron emission yield as a function of impinging electron energy for dust particles of two materials (\textit{red} -- astronomical silicates, \textit{blue} -- carbonaceous) and different sizes (\textit{dashed} -- $10$ nm, \textit{solid} -- $100$ nm, \textit{dash-dotted} -- $1000$ nm). Reproduced by the data from \cite{kimura_mann_1999}.}
    \label{figure: secondary_emission_yield}
\end{figure}

Collisional charging leads to negatively charged dust grains because, for the same temperature, the electrons have larger thermal velocities than protons and, therefore, higher fluxes towards the grain surface (see, e.g., formula (\ref{formula: reduced_sticking_rate})). At the same time, high-energy electrons and protons falling on the surface of a dust grain can excite electrons at this surface and produce the current of secondary electrons emitting from the grain. Thus, this current increases the charge of a dust grain, especially within the hot plasma environment. A key parameter describing the resulting current is the secondary electron emission yield $\delta_{i}$, which is the number of secondary electrons produced per impact (Figure \ref{figure: secondary_emission_yield}). The value $\delta_{i}$ depends on the dust grain properties and the initial energy of impinging plasma particles. Here, we use the values $\delta_{i}$ from \cite{kimura_mann_1999}.

One can express the current of secondary electrons in terms of the number of collisions between plasma particles and a dust grain, similar to that in Section \ref{subsection: sticking}. However, one should also consider (a) the secondary electron yield $\delta_i$ and (b) the fact that only those secondary electrons, which have energy greater than $\varepsilon_* = max(0, eU)$, can leave the surface of a dust grain. The last fact could be taken into account by a correction factor obtained from the energy probability distribution of secondary electrons $\rho_i$ \citep[for details, see][]{kimura_mann_1998}, where index $i$ denotes the sort of particles, which induced the emission of secondary electrons ($i$ is either $p$ or $el$). Therefore, the current of secondary electrons produced by plasma particles of sort $i$ is:

\begin{equation} \label{formula: secondary_emission_current}
    J^{see}_i = e     
    \int\limits_{0}^{\infty} \int\limits_{0}^{\pi}\int\limits_{0}^{2\pi}\delta_{i}(E)\sigma_{i}(v)f_{i}(v, \vartheta)v^3\sin{\vartheta}d{\varphi}d{\vartheta}dv \cdot \int\limits_{\varepsilon_{*}}^{\infty}\rho_i(\varepsilon)d\varepsilon,
\end{equation}

\noindent
where $E = \frac{m_iv^2}{2} + q_ieU$ is the initial energy of impinging plasma particle. Here, we consider secondary electron emission only induced by plasma electrons since the current corresponding to plasma protons is significantly lower than the one corresponding to plasma electrons because of the big difference in mass of these particles.

One can reduce the dimensionality of integral \ref{formula: secondary_emission_current} in the same way as was performed for the corresponding integral in \ref{appendix: reduced_collisional_integrals} to increase the computational efficiency.

\subsection{Photoemission}
\label{subsection: photoelectric emission}

The surroundings of ISD grains have a strong UV radiation field governed by solar and interstellar background photons. As a result, several mechanisms which generate currents of electrons from the surface of a dust grain appear. Here we follow the approach proposed in \cite{weingartner_2006}. They considered four different types of currents driven by: (1) primary photoelectrons (the electrons excited directly due to the absorption of photons), (2) Auger electrons, (3) secondary photoelectrons, and (4) photodetachment. Each of these currents depends on the photon flux, cross section, threshold energy, and yield corresponding to the process.

\begin{figure}
    \center{\includegraphics[width=1\linewidth]{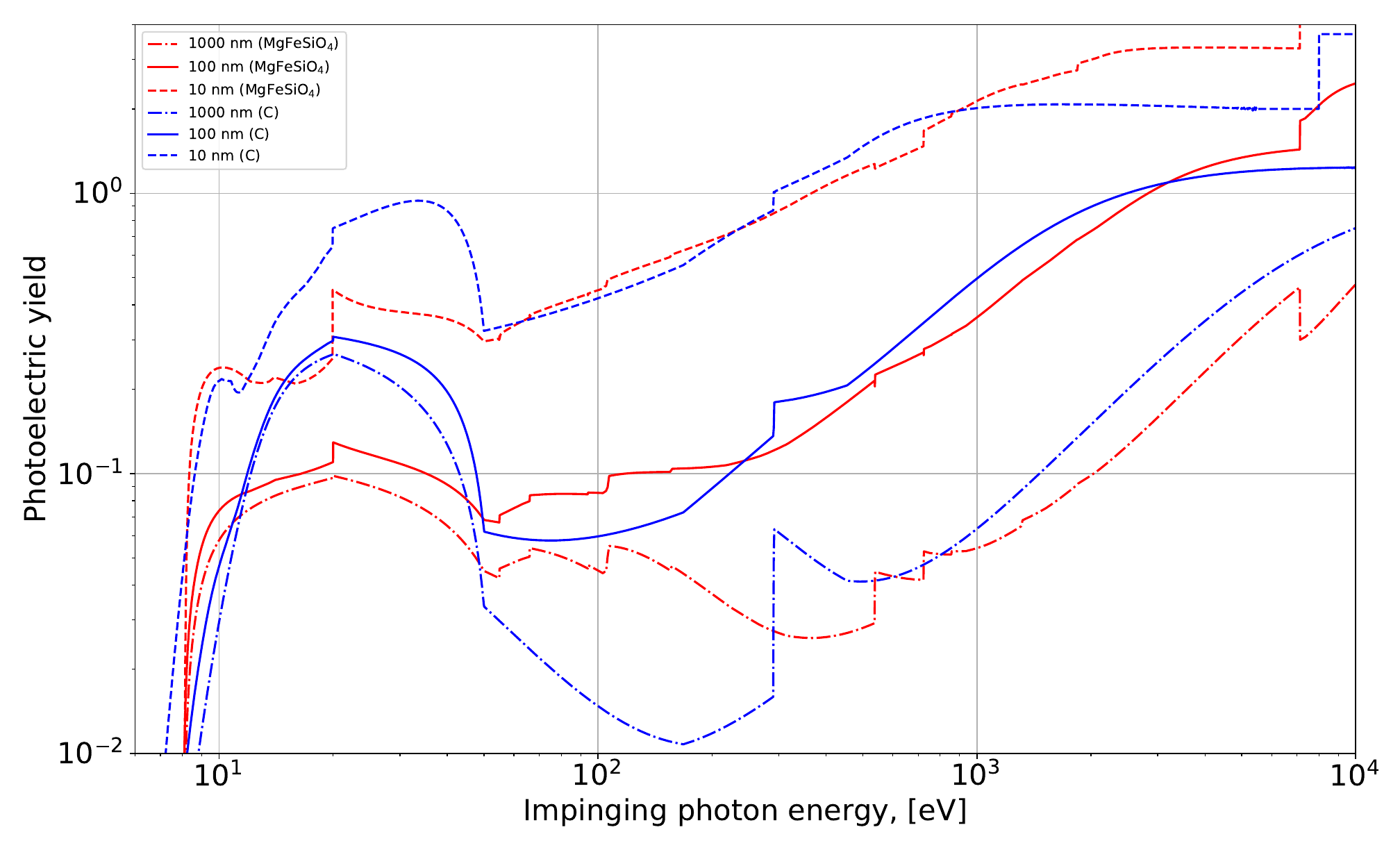}}
    \caption{The total photoelectric yield as a function of impinging photon energy for dust particles of two materials (\textit{red} -- astronomical silicates, \textit{blue} -- carbonaceous) and different sizes (\textit{dashed} -- $10$ nm, \textit{solid} -- $100$ nm, \textit{dash-dotted} -- $1000$ nm). Computed using the approach from \cite{weingartner_2006}.}
    \label{figure: photoelectric_yield}
\end{figure}

It is convenient to join the first three types of photoelectric currents into one group:

\begin{equation*}
    J^{pe} = e \int\limits_{E_{pet}}^{E_{max}}F(E)C_{abs}(E)Y(E)dE,
\end{equation*}

\noindent
where $E_{pet}$ is the threshold energy for the emission of photoelectrons, $E_{max}$ is the maximum photon energy in the incident radiation field, $F$ is the photon flux (see \ref{appendix: photoemission}), $C_{abs}$ is the absorption cross section, $Y$ is the total photoelectric yield (Figure \ref{figure: photoelectric_yield}), i.e., the number of photoelectrons emitting from a dust grain per absorbed photon, which consists of three components:

\begin{equation*}
    Y(E) = Y^{p}(E) + Y^{a}(E) + Y^{s}(E),
\end{equation*}

\noindent
where $Y^{p}$ is the primary photoelectrons yield, $Y^{a}$ is the yield of Auger electrons, $Y^{s}$ is the secondary photoelectrons yield. One can find out all details and exact expressions for these values in \cite{weingartner_2006}. The absorption cross-section $C_{abs}$ is obtained from the Mie theory, and optical constants are taken from \cite{draine_2003} \footnote{\href{https://www.astro.princeton.edu/~draine/dust/dust.diel.html}{Optical constants for smoothed UV astronomical silicates and carbonaceous grains.}}

The last current corresponding to photoelectric emission is the photodetachment current:

\begin{equation*}
    J^{pd} = e \int\limits_{E_{pdt}}^{E_{max}}F(E)\sigma_{pdt}(E)dE,
\end{equation*}

\noindent
where $E_{pdt}$ is the photodetachment threshold energy, $\sigma_{pdt}$ is the photodetachment cross section, and it is implicitly assumed that $Y^{pdt} = 1$. All details and formulae for these values can be found in \cite{weingartner_2001}.

\subsection{Cosmic ray effects}
\label{subsection: cr_effects}

For the first time, we demonstrate the effects of cosmic rays on ISD charging in the LISM. In this Section, we fully follow the theory developed by \cite{ivlev_2015} and its implementation by \cite{ibanez_2019}.

In the interstellar space Voyager 1 registered the increased fluxes of high-energy cosmic rays \citep{stone_2013}. \cite{ivlev_2015} showed that within certain surrounding conditions in the interstellar medium, cosmic rays influence dust charging significantly, which is why, in general, the effects of these particles should be considered apart from the thermal plasma component.

The spectrum of cosmic rays could be extrapolated from Voyager 1 measurements \citep{ivlev_2015}:

\begin{equation} \label{formula: cosmic_ray_flux}
    j_k (E) = C_k \frac{E^{\alpha_k}}{(E + E_0)^{\beta_k}} \:\: \text{eV}^{-1} \text{cm}^{-2} \text{s}^{-1} \text{sr}^{-1},
\end{equation}

\noindent
where $k$ is a sort of cosmic ray particles (protons or electrons), $E_0$ = 500 MeV. Table \ref{table: cosmic_ray_flux_parameters} presents the numerical values of parameters appearing in equation (\ref{formula: cosmic_ray_flux}). It is seen that the flux of cosmic ray electrons is higher than the flux of protons, which is why we neglect the influence of cosmic ray protons in our computations. Thus, the current corresponding to the combined collection of cosmic ray electrons onto ISD grains and secondary electron emission induced by them is as follows \citep{ivlev_2015, ibanez_2019}:

\begin{equation} \label{formula: cosmic_ray_current}
    J^{cr} = -e \pi a^2 \int\limits_{E_{int}}^{\infty}4\pi j_{el}(E)[s^{cr}_{el}(E) - \delta_{el}(E)]dE,
\end{equation}

\noindent
where $E_{int} \approx 1.5 \times 10^{-2} \: \text{eV}$ is the intersection energy, i.e., the value for which the flux of cosmic ray electrons equals the flux of thermal electrons, $s_{el}^{CR}(E)$ is the sticking coefficient for cosmic ray electrons \citep[see, e.g., Appendix C in][]{ivlev_2015}. We use another expression for sticking coefficient (as compared with the one in \ref{appendix: sticking_coefficient}) to take account of the fact that high-energy electrons pass through the grain and, therefore, do not influence the collection part of the current (\ref{formula: cosmic_ray_current}). Thus, the corresponding boundary for astronomical silicates of radius $a = 100$ nm is $\sim$ 3 keV. In Section \ref{subsection: results_steady_state}, we discuss how taking into account the cosmic ray effects influences the values of ISD steady-state potential.

In this study, we do not consider the effects of other energetic plasma components, e.g., anomalous cosmic rays or solar energetic particles, because we expect they do not influence the ISD charging significantly. Anomalous cosmic rays are the interstellar pickup-ions accelerated at the Termination Shock from energies of $\sim$ 1 keV (freshly ionized pickup-ion) up to the energies of tens of MeV \citep[see, e.g., ][]{giacalone_2022}. For such high-energy particles, the value of the sticking coefficient equals 0. The lower end of the energy spectrum of solar energetic particles ($\sim$ tens of keV), for which the value of the sticking coefficient is non-zero, overlaps with the high energy particles of solar wind plasma. That is why the influence of this part of solar energetic particles on ISD charging is negligible compared with the sticking of primary plasma particles. At the same time, the principal part of the solar energetic particles spectrum is in the range \citep[see, e.g., ][]{mcguire_1984} where the corresponding value of the sticking coefficient equals 0. Thus, there is no need to consider anomalous cosmic rays or solar energetic particles as individual components when probing the ISD charging.

\begin{table}
    \centering
    \caption{Parameters of cosmic ray spectra.}
    \begin{tabular}{|l|l|l|l|}
    \hline
    Sort (k) & $C_k$ & $\alpha_k$ &  $\beta_k$ \\
    \hline
    Electrons & 2.1 $\times$ 10$^{18}$ & -1.5        & 1.7 \\
    \hline
    Protons & 2.4 $\times$ 10$^{15}$ & -0.8        & 1.9 \\
    \hline
    \end{tabular}
    \label{table: cosmic_ray_flux_parameters}
\end{table}

\section{Results} \label{section: results}

The coordinate system used to demonstrate the modeling results comes from the global heliospheric model of \cite{izmodenov_alexashov_2020}: $Z$-axis is directed toward the interstellar flow (upwind direction), $X$-axis is in the BV-plane (plane containing the velocity ${\bf V}_{LISM}$ and magnetic field induction ${\bf B}_{LISM}$ vectors of the undisturbed LISM) and perpendicular to $Z$-axis. The direction of the $X$-axis is chosen such that the projection of ${\bf B}_{LISM}$ to the $X$-axis is negative. $Y$-axis completes the right-handed system. The direction of ${\bf V}_{LISM}$ is (longitude = 75.4$^{\circ}$, latitude = -5.2$^{\circ}$) in ecliptic (J2000) coordinate system. The direction of ${\bf B}_{LISM}$ is (longitude = 125$^{\circ}$, latitude = 37$^{\circ}$) in heliographic inertial (HGI) coordinate system.

\subsection{Currents} \label{subsection: results_currents}

\begin{figure}
    \center{\includegraphics[width=1\linewidth]{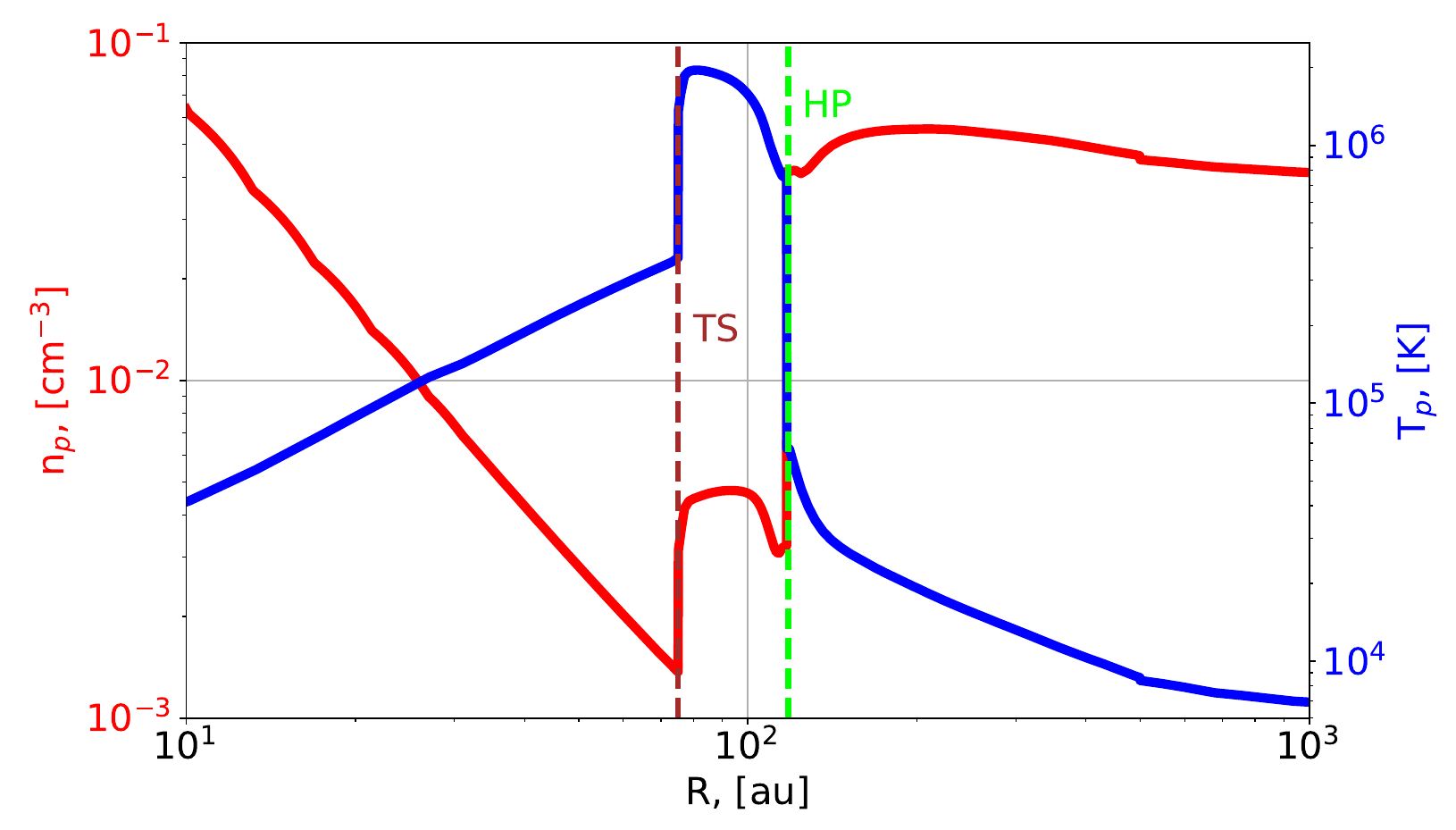}}
    \caption{The distributions of plasma parameters along the upwind direction obtained from the heliospheric model of \protect\cite{izmodenov_alexashov_2020}. Red and blue lines correspond to proton/electron number density and temperature, respectively (the magnitudes for protons and electrons are the same in the heliospheric model used). Brown and green vertical dashed lines correspond to the heliospheric termination shock (TS, 75 au) and the heliopause (HP, 117 au). Heliospheric discontinuities separate the regions with essentially different conditions for dust charging.}
    \label{figure: plasma_distributions}
\end{figure}

In this Section, we present the results of computations for currents $J_k$ produced by different physical processes described in Section \ref{section: currents}. The currents depend strongly on the local plasma and radiation conditions. Figure \ref{figure: plasma_distributions} presents the distributions of proton/electron number density and temperature along the upwind direction obtained in the frame of \cite{izmodenov_alexashov_2020} model. Both density and temperature are essentially different in the supersonic solar wind, in the inner heliosheath (between the TS and HP), and in the LISM. 

\begin{figure*}
    \center{\includegraphics[width=1\linewidth]{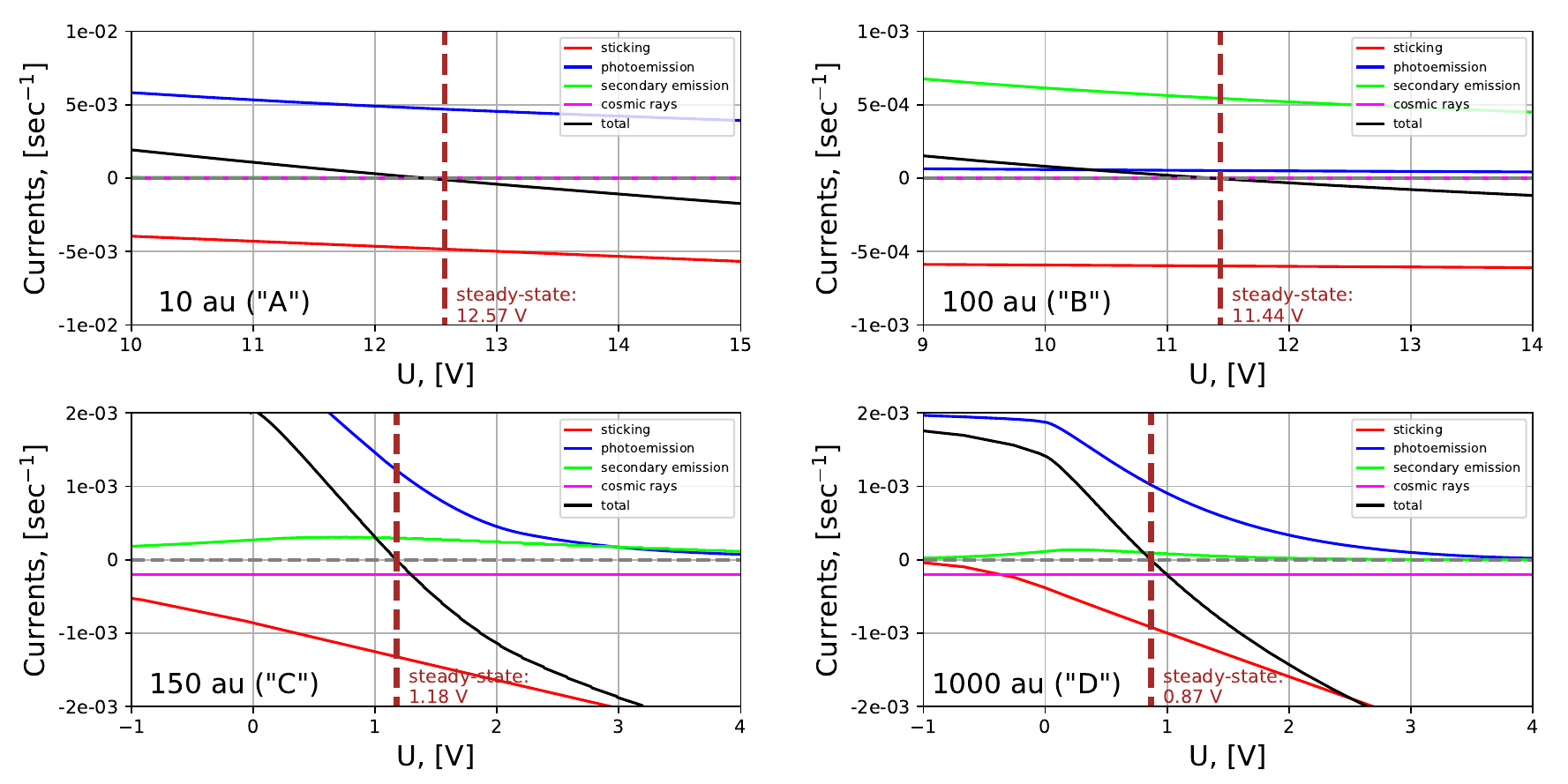}}
    \caption{Currents as functions of surface charge potential for astronomical silicates of radius $a = 100$ nm. Each panel corresponds to the particular point located at the specified heliodistance along the upwind direction (points "A" -- "D", see definitions in the text). Different colors match different groups of physical processes. Black curve corresponds to the total current, i.e., sum over all groups. Grey dashed horizontal line highlights identically zero current for visibility. Brown dashed vertical line is the value of steady-state potential. All currents are divided by the absolute measure of the elementary charge $e$.}
    \label{figure: heliospheric_currents}
\end{figure*}

Figure \ref{figure: heliospheric_currents} presents currents as functions of surface charge potential for astronomical silicates of size $a = 100$ nm at four particular points in the heliospheric interface. The points "A", "B", "C", "D" are located at $Z$-axis in the upwind direction at 10, 100, 150, 1000  au, respectively. The four points correspond to four different regions of the heliospheric interface: point "A" is in the supersonic solar wind, point "B" is in the inner heliosheath between TS and HP, point "C" is in the perturbed interstellar medium, and point "D" is in the pristine interstellar medium.

It is seen from Figure \ref{figure: heliospheric_currents} that, in all regions, the dominant negative current is formed by the sticking of thermal plasma electrons. Cosmic ray electrons have a small additional impact in the LISM: 10-20 \% of the current of thermal electrons (panels "C" and "D" from Figure \ref{figure: heliospheric_currents}). The nature of the dominant positive current depends on the region of the heliospheric interface. There are essentially just two options for positive currents -- photoemission and secondary electron emission since the sticking of thermal plasma protons is negligible compared to these two processes.

Let us start with point "A". In the inner heliosphere at small heliocentric distances, plasma is relatively cold (Figure \ref{figure: plasma_distributions}). At the same time, the influence of solar radiation is significant as it is inversely proportional to the squared heliodistance. That is why in the inner heliosphere, the dominant positive current is photoemission (Figure \ref{figure: heliospheric_currents}, panel "A"). In the inner heliosheath, where point "B" is located, plasma temperature rapidly increases, which leads to the enhancement of secondary electron emission production rate -- more energetic thermal electrons produce more secondary electrons emitting from the grain. At the same time, the flux of solar photons decreases, and the contribution of interstellar background photons becomes of the order of solar photons (see Figure \ref{figure: total_photon_flux_spectrum}). Therefore, total photon flux also decreases as compared with its value at point "A" and, thus, the dominant positive current in the inner heliosheath is secondary electron emission (Figure \ref{figure: heliospheric_currents}, panel "B"). The perturbed interstellar medium is characterized by decreasing plasma temperature and approximately constant flux of photons because the principal part of the total photon spectrum (low-energy photons) is due to the interstellar background photons. It means that the importance of secondary electron emission is decreasing while the rate of photoemission keeps its value. Thus, near the heliopause, where point "C" is located, both photoemission and secondary electron emission are essential components of total positive current (Figure \ref{figure: heliospheric_currents}, panel "C"), while in the pristine interstellar medium (point "D") the secondary electron emission becomes negligible compared to the photoemission (Figure \ref{figure: heliospheric_currents}, panel "D").

\subsection{Steady-state potential} \label{subsection: results_steady_state}

By definition, the steady-state potential is the potential for which the total current equals zero. In Figure \ref{figure: heliospheric_currents} the values of steady-state potential are highlighted by brown vertical dashed lines. Inside the heliosphere, absolute values of steady-state potential are higher than in the LISM because of more intensive rates of secondary electron emission and photoemission. These rates are induced by relatively high plasma temperature (Figure \ref{figure: plasma_distributions}) and heavy photon fluxes (Figure \ref{figure: total_photon_flux_spectrum}), respectively.

\begin{figure}
    \center{\includegraphics[width=1\linewidth]{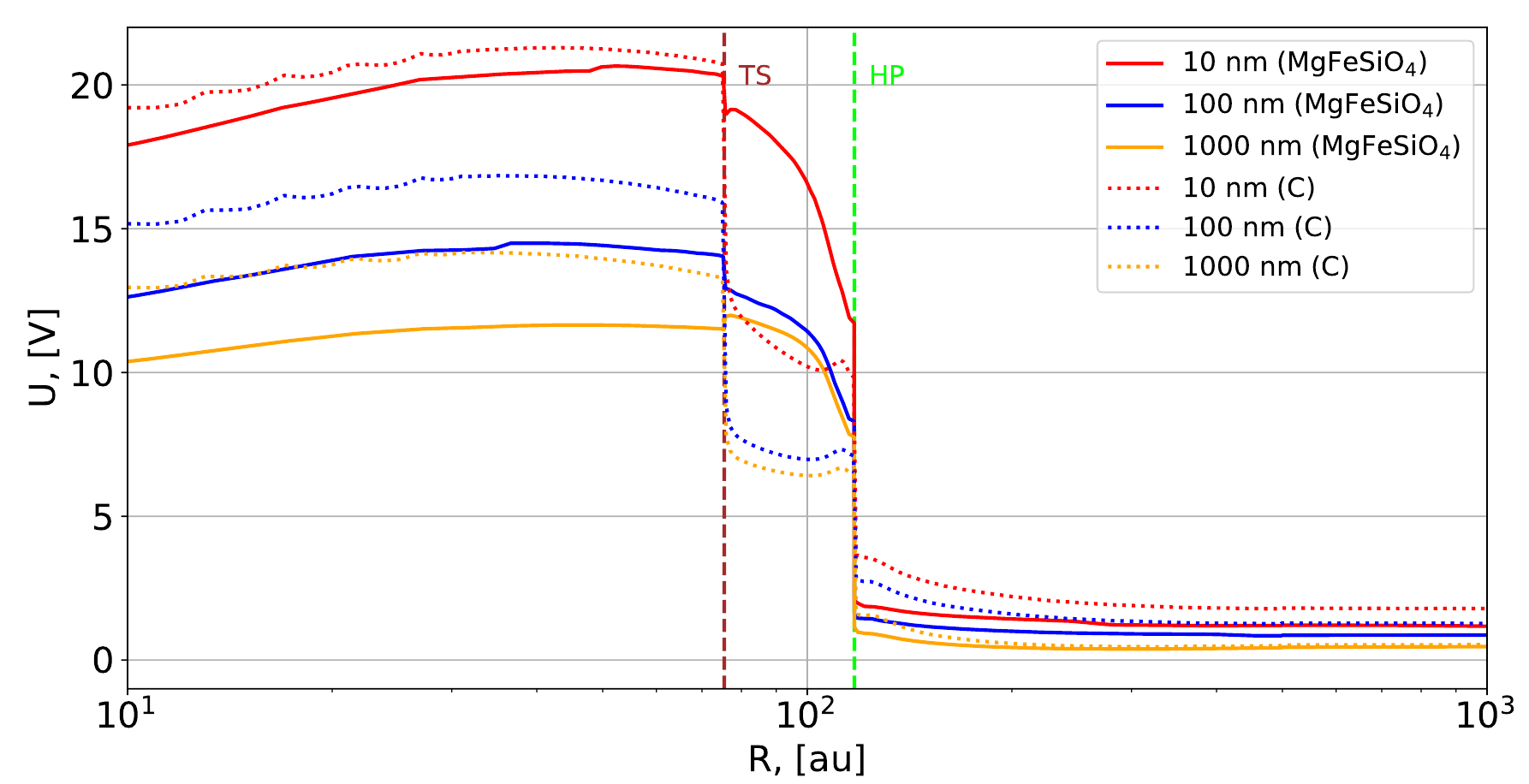}}
    \caption{The distributions of steady-state potential along the upwind direction for ISD particles of three sizes: $a = 10$ (red), $100$ (blue), $1000$ (orange) nm, and two materials: astronomical silicates (solid) and carbonaceous grains (dotted). Brown and green vertical dashed lines correspond to the heliospheric termination shock and the heliopause.}
    \label{figure: potentials}
\end{figure}

In Figure \ref{figure: potentials} the distributions of steady-state potential along the upwind direction are presented for dust grains of three sizes ($a = 10, 100, 1000$ nm) and two materials (astronomical silicates and carbonaceous grains).

For all considered types of grains, dependencies of steady-state potential on the heliocentric distance are qualitatively similar. In the LISM, the value of steady-state potential is approximately constant because plasma parameters and interstellar background radiation do not change significantly. Across the transition from the LISM into the heliosphere, the biggest jumps of steady-state potential appear, which is why, in this region, the validity of steady-state assumption must be verified (Section \ref{section: dynamical_effects}). Up to the crossing of the termination shock, the steady-state potential rises because of plasma temperature increase. Inside the termination shock, the value of steady-state potential is almost constant again. The reason is that the rates of photoemission and sticking of primary electrons are approximately inversely proportional to the squared heliodistance. The assumption of constant charge potential inside the region of supersonic solar wind was used for modeling by many authors \citep[see, e.g.,][]{landgraf_2000, sterken_2012, sterken_2013, strub_2019, mishchenko_2020, godenko_2021a}.

The photoemission is slightly more effective for carbonaceous grains than for astronomical silicates (Figure \ref{figure: photoelectric_yield}), while with the secondary electron emission the situation is opposite (Figure \ref{figure: secondary_emission_yield}). As a result, the values of steady-state potential in the supersonic solar wind and in the LISM, where the photoemission is a dominant positive current (Section \ref{subsection: results_currents}) are higher for carbonaceous grains than for astronomical silicates. Between the TS and HP, in contrast, the values for astronomical silicates are higher because, in this region, the secondary electron emission plays a key role.

One should also note that, for the smallest considered particles of radius $a = 10$ nm, in the LISM, the absolute value of charge is relatively small ($Z = Q/e \approx $ 10-15), and, therefore, the kinetic approach may be required in order to probe the ISD charging accurately. That is why in \ref{appendix: stochastic_charging}, we show the computational results of equilibrium stochastic charge distribution within different heliospheric conditions.

It is seen from Figure \ref{figure: potentials} that the value of steady-state potential is higher for smaller dust grains. The reason is that the photoelectric and secondary electron emission yields (Section \ref{section: currents}) depend on the curvature of grain's surface. For smaller spherical particles, the values of yield are larger because, in this case, emitted electrons need less time to leave the surface of a dust grain, and that is why the probability of their absorption along the path to the surface is less as well \citep{kimura_mann_1999}.

It should be also noted that for bigger dust grains, the dependence of steady-state potential on size is weaker \citep[see, e.g., Figure 2 from][]{slavin_2012}. The reason is that, as seen for protons and electrons, the surface of big dust grains looks like a plane. That is why, in this case, the values of photoelectric and secondary electron emission yields are close to the corresponding values for semi-infinite slab \citep{kimura_mann_1999}, and their dependence on size disappears.

As was mentioned above, in this study, we, for the first time, attempt to consider the effects of cosmic rays on ISD charging. Table \ref{table: cr_effects} presents the values of steady-state potential computed with and without taking into account the cosmic ray effects in the pristine LISM at the heliocentric distance of 1000 au. For all considered grain sizes, the absolute difference of the values of steady-state potentials between two cases is approximately 0.1-0.15 V. The biggest absolute and relative difference is observed for large ISD particles. However, the dynamics of large grains is less influenced by the electromagnetic force (the magnitude is inversely proportional to $a^{2}$), and, therefore, they are less affected by small changes of steady-state potential. At the same time, for small ISD particles, the relative difference does not exceed 10 \%, which is small enough. Thus, the effects of cosmic rays on ISD charging are visible, but not crucial.

\begin{table}
    \centering
    \caption{The influence of cosmic rays on the value of steady-state potential in the pristine LISM (1000 au) for astronomical silicates.}
    \begin{tabular}{|l|l|l|}
    \hline
    Size    & With CR, [V] & Without CR, [V] \\
    \hline
    $a = 10$ nm & 1.19  & 1.29 \\
    \hline
    $a = 100$ nm & 0.87 & 1.00 \\
    \hline
    $a = 1000$ nm & 0.47 & 0.64 \\
    \hline
    \end{tabular}
    \label{table: cr_effects}
\end{table}

\subsection{Comparison with previous charging models}
\label{subsection: comparison_models}

The values of steady-state potential computed by our charging model are higher than the values obtained by previous modelers within the heliospheric conditions \citep[see, e.g.,][]{kimura_mann_1998, slavin_2012, alexashov_2016}. However, there are two sources for the origin of discrepancies between the results obtained by different charging models: (1) formulae for computation of currents and (2) distributions of plasma parameters. The formulae for currents applied in the charging model from \cite{slavin_2012} are similar to those used in our model. \cite{slavin_2012} used the heliospheric model from \cite{pogorelov_2008}, and the boundary conditions for plasma parameters were the following: in the LISM at 1200 au: $n = 0.06$ cm$^{-3}$, $v = 26.4$ km s$^{-1}$, $T = 6527$ K, and in the solar wind at 10 au: $n = 0.07$ cm$^{-3}$, $v = 454$ km s$^{-1}$, $T = 3800$ K.

In Table \ref{table: comparison} the values of steady-state potential obtained from our and Slavin’s charging models at two mentioned points for astronomical silicates of different sizes are presented. One can see that in the LISM, the differences between the corresponding values are negligible, while in the inner heliosphere, the difference of 1-1.5 V appears. This discrepancy comes from slightly different formulae used for the rates of sticking and secondary emitted electrons. However, this discrepancy is not critical and does not influence the analysis of dynamical charging of ISD grains presented in Section \ref{section: dynamical_effects}.

To assess the influence of different plasma distributions on the value of steady-state potential, we compare the values for grains of radius $a = 100$ nm from Table \ref{table: comparison} and the values from Figure \ref{figure: heliospheric_currents} (Panels "A" and "D"). The resulting relative difference in the inner heliosphere is approximately 30 \%, while in the pristine LISM, it is 20 \%. The future reliable measurements of ISD charge (onboard, e.g., IMAP or Interstellar Probe) could decrease this relatively high level of uncertainty.

\begin{table*}
    \centering
    \caption{Comparison between the values of steady-state potential for astronomical silicates obtained from different charging models}
    \begin{tabular}{|l|l|l|l|}
    \hline
    Model    & $a = 10$ nm, [V] & $a = 100$ nm, [V] & $a = 1000$ nm, [V] \\
    \hline
    Our, LISM & 0.96        & 0.69        & 0.31\\
    \hline
    \cite{slavin_2012}, LISM & 0.98        & 0.67        & 0.33\\
    \hline
    Our, solar wind & 12.18        & 8.32        & 6.84\\
    \hline
    \cite{slavin_2012} , solar wind & 10.92        & 6.98        & 5.61\\
    \hline
    \end{tabular}
    \label{table: comparison}
\end{table*}

\section{Effects of dynamically changed charge} \label{section: dynamical_effects}

In this Section, we show how ISD charge modifies across the heliopause according to equation (\ref{formula: general_equation}). We also demonstrate how the dynamically changed charge influences the density distributions of ISD grains compared to the case of steady-state approximation. Here, we consider only the effects on the astronomical silicates.

\subsection{Relaxation time across the heliopause}
\label{subsection: relaxation_time}

Let us consider how ISD grains pass through the heliopause because, across this surface, the biggest jumps of steady-state charge potential appear. We explore how the value of ISD charge potential approaches the magnitude of steady-state potential for particles of different sizes. For this purpose, we study a test case where the dynamics of ISD is affected just by the electromagnetic force (at large heliocentric distances, the influence of gravitation and radiation pressure is negligible) and, therefore, the system of motion equations is as follows:

\begin{equation} \label{formula: isd_dynamics}
    \begin{cases}
        \frac{d \vec{r}}{dt} = \vec{v}, \\
        \frac{d \vec{v}}{dt} = \frac{Q}{m} \left[(\vec{v} - \vec{v}_p) \times \vec{B}\right],
    \end{cases}
\end{equation}

\noindent
where $\vec{v}_p$ is the bulk velocity of plasma and $\vec{B}$ is the magnetic field induction vector. To demonstrate the effect, we also assume that ISD grains start their motion in front of the heliopause with an initial velocity of undisturbed LISM (the magnitude is 26.4 km s$^{-1}$ relative to the Sun and directed anti-parallel to the upwind direction). The grains cross the heliopause and penetrate the heliosphere, where the plasma conditions change, as well as the value of steady-state potential.

\begin{figure}
    \center{\includegraphics[width=1\linewidth]{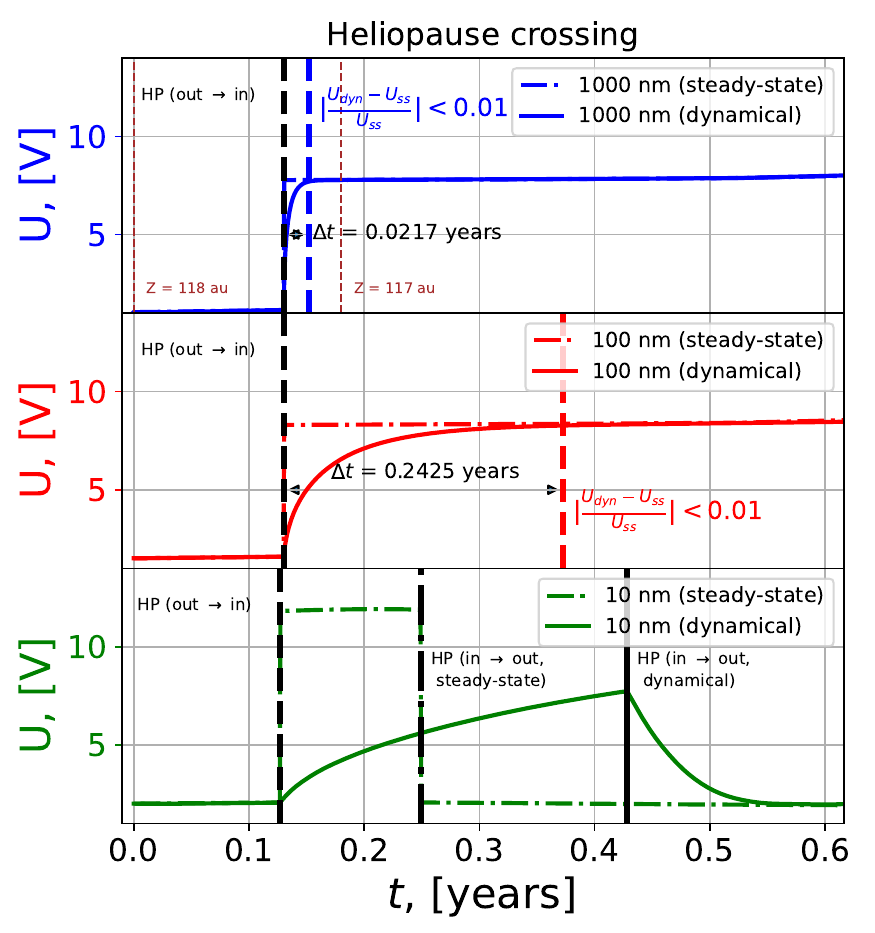}}
    \caption{The evolution of dust grain charge during the passage through the heliopause for astronomical silicates. The top, middle and bottom panels correspond to big ($a$ = 1000 nm), medium-sized ($a$ = 100 nm) and small ($a$ = 10 nm) grains, respectively. The results obtained with dynamical charge computation and steady-state charge approximation are denoted by solid (joint solving of (\ref{formula: isd_dynamics}) and (\ref{formula: general_equation})) and dash-dotted (solving of (\ref{formula: isd_dynamics}) using the steady-state charge approximation) lines, correspondingly. The solver of systems of ordinary differential equations is a fourth-order Runge-Kutta method. Black dashed vertical lines correspond to the moments of time when dust particles pass through the heliopause. $\Delta t$ is the relaxation time, i.e., the time required for the grain to reach the state with relative deviation of charge $\Delta \hat{Q}$ less than $1 \%$. Brown dashed vertical lines in the top panel indicate the moments of intersection with the corresponding planes.}
    \label{figure: heliopause_crossing}
\end{figure}

In Figure \ref{figure: heliopause_crossing} the analysis of how the relaxation of charge takes place across the heliopause is presented for ISD particles of three sizes (1000 nm -- \textit{big} grains, 100 nm -- \textit{medium-sized} grains, 10 nm -- \textit{small} grains) via the evolution of surface charge potential along the trajectory. By vertical black dashed lines, we denote the moments when ISD grains pass through the heliopause. The moments are roughly the same for all considered types of grains. We assume that a dust grain reaches an equilibrium if the following condition is satisfied:

\begin{equation} \label{formulae: criterion}
    \Delta \hat{Q} = \bigg| \frac{Q - Q_{eq}}{Q_{eq}} \bigg| < 10^{-2},
\end{equation}

\noindent
where $Q$ is the actual charge of a dust grain obtained from the solving of differential equation (\ref{formula: general_equation}), $Q_{eq}$ is the local steady-state charge. This criterion is arbitrary (we can choose another number instead of $10^{-2}$), but it allows estimating the dependence of relaxation time on radius. By colored vertical dashed lines, we denote the moments when the criterion becomes valid. The time difference $\Delta t$ between colored and corresponding black lines is the proper estimate of the time required for grains to reach an equilibrium when crossing the heliopause. As seen from Figure \ref{figure: heliopause_crossing}, for big ISD grains, the relaxation time is negligible ($\sim$ 0.02 years) compared to the time scale of ISD motion in the heliosphere ($\sim$ 0.2 years to cover the distance of 1 au). Therefore, the accurate dynamical computation of charge is redundant in this case. For medium-sized ISD grains, the relaxation time scale increases ($\sim$ 0.25 years) and becomes comparable with the characteristic time scales of ISD motion in the heliosphere. The increase in relaxation time is due to the fact that absolute values of all currents considered in our charging model are approximately proportional to $\pi a^2$ (see Section \ref{section: currents}) and, thus, for bigger grains, charging proceeds faster ($dt \approx dQ / J$). One should also note that, as seen in Figure \ref{figure: heliopause_crossing}, the relaxation time is roughly inversely proportional to the radius of a dust grain. The approximate estimate of relaxation time also reproduces this trend (see, for details, \ref{appendix: rough_estimate}). For small grains, there is no corresponding colored vertical dashed line in Figure \ref{figure: heliopause_crossing} because these particles are swept away by the solar wind and go back into LISM. The period of time inside the heliosphere for these grains is insufficient to satisfy the introduced criterion (\ref{formulae: criterion}). In this case, the moments when ISD grains go out from the heliosphere differ significantly depending on the approach for the computation of charge (black solid and dash-dotted lines), which is why trajectories depend on it as well.

The analysis shows that, for big ISD grains, a steady-state charge assumption is valid and should be used due to its computational efficiency. In case of medium-sized and small grains we expect that some effects on density distribution could appear, and a more rigorous analysis is required to check this hypothesis.

\subsection{Influence on density distribution}
\label{subsection: density_distributions}

In this Section, we show how the dynamical computation of charge influences the ISD density distribution in the heliosphere compared to the use of steady-state approximation. To calculate ISD number density, we use the Monte Carlo method \cite[see, e.g.,][]{godenko_2021a}. In this Section, we present the results for dust particles of two sizes: $a$ = 100 nm and $a$ = 10 nm, because for big ISD particles of size $a$ = 1000 nm, the difference between the results is negligible. We study the effect near the heliopause, where the largest changes of surrounding plasma conditions appear.

\begin{figure*}
    \center{\includegraphics[width=1\linewidth]{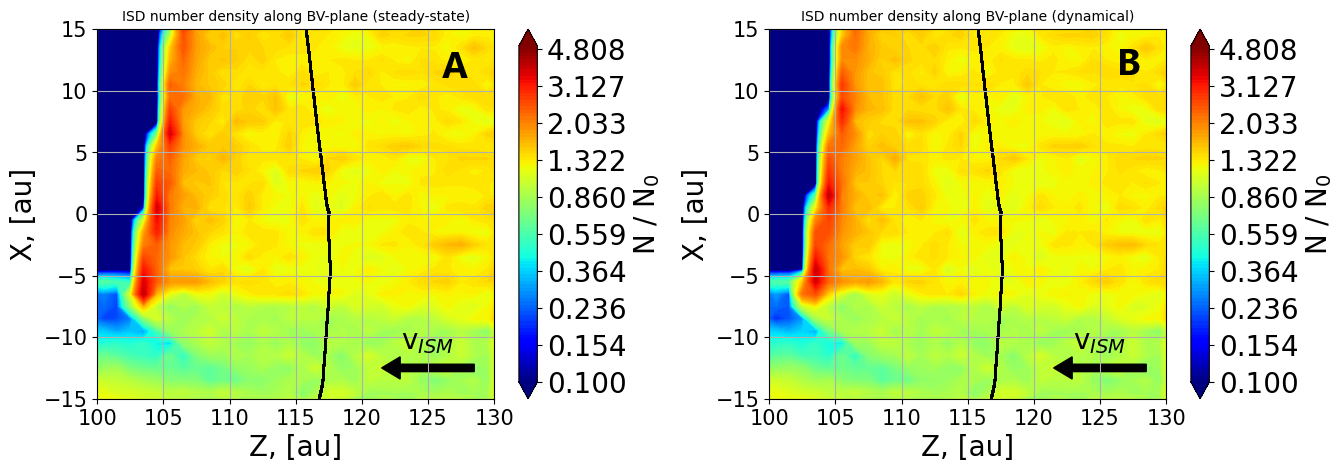}}
    \center{\includegraphics[width=1\linewidth]{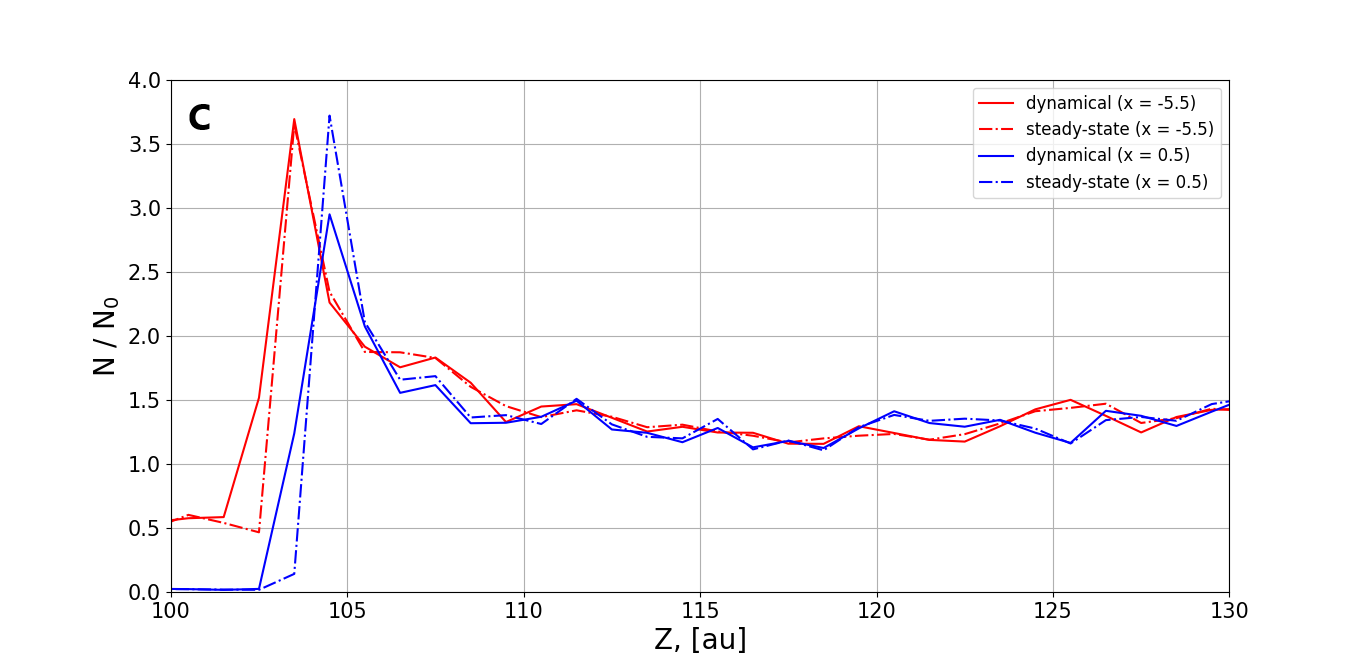}}
    \caption{The ISD number density distributions (normalized to the magnitude of number density $N_0$ in the undisturbed LISM) along BV-plane in the nose part of the heliosphere obtained by Monte Carlo computations with the use of steady-state potential approach (Panel A) and with dynamically changed charge (Panel B) for astronomical silicates of size $a$ = 100 nm. The sizes of computational cells are 1 au $\times$ 1 au $\times$ 1 au. The magnitude or relative statistical error is less than 5$\%$ for any point. Black curve matches the heliopause. Panel C represents the 1D distributions of ISD number density along lines $X = -5.5$ and $X = 0.5$ for both approaches of charge computation. Solid and dash-dotted lines denote dynamically changed and steady-state charges, correspondingly.}
    \label{figure: density_distributions_a=100nm}
\end{figure*}

In Figure \ref{figure: density_distributions_a=100nm} the results of Monte Carlo modeling of ISD number density are presented for particles of size $a$ = 100 nm. The effect is almost invisible from 2D number density maps (panels A and B). Looking at panel C, where the 1D distributions along the lines $X = const$ are demonstrated, one can notice that trajectories of ISD particles in case of dynamically changed charge go slightly deeper inside the heliosphere. Moreover, the magnitude of ISD number density at the region of dust accumulations decreases (blue lines in panel C). However, the effect is relatively small, and a steady-state charge approximation for these particles is quite justified.

\begin{figure*}
    \center{\includegraphics[width=1\linewidth]{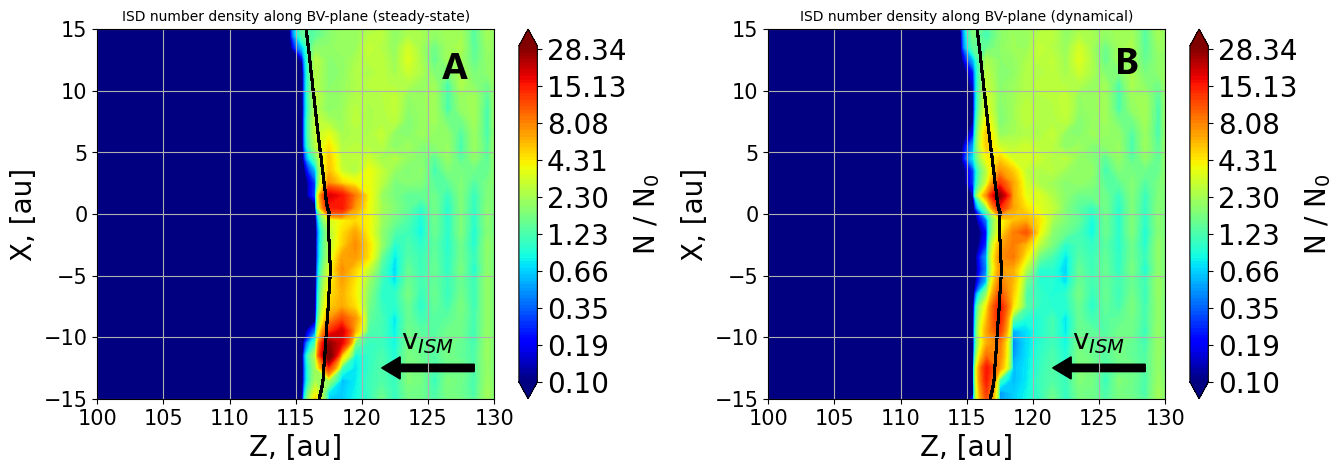}}
    \center{\includegraphics[width=1\linewidth]{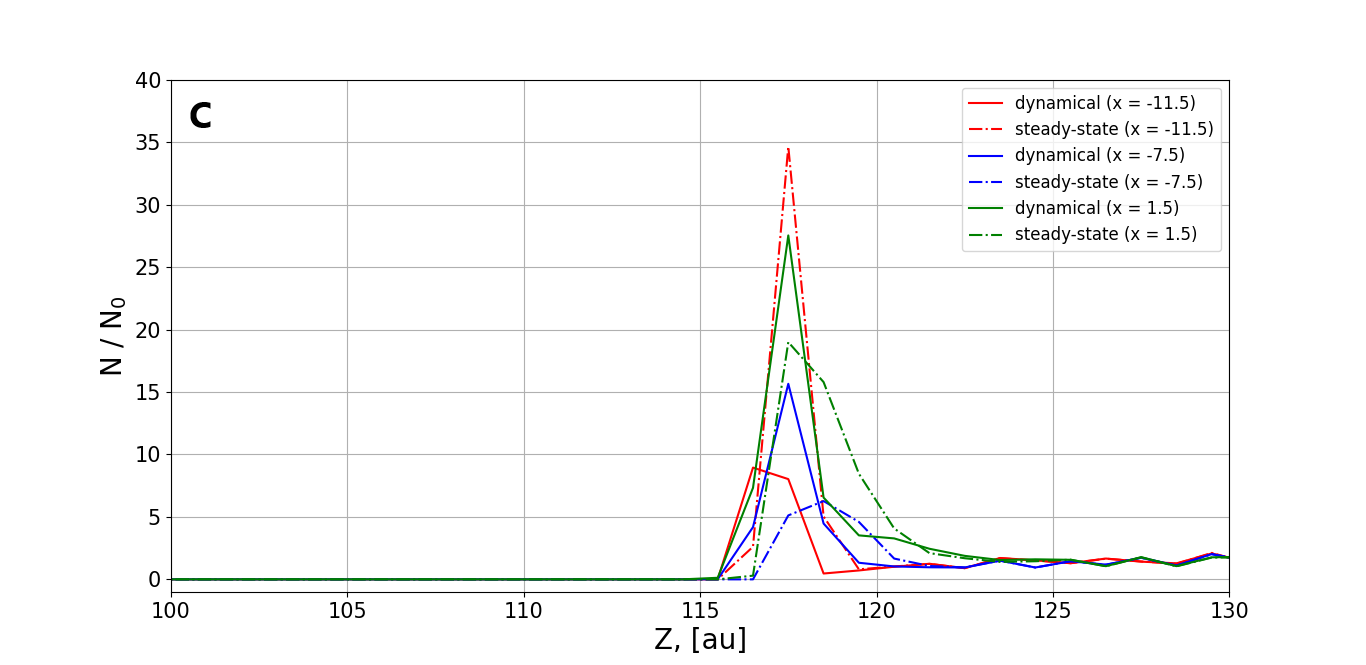}}
    \caption{The ISD number density distributions (normalized to the magnitude of number density $N_0$ in the undisturbed LISM) along BV-plane in the nose part of the heliosphere obtained by Monte Carlo computations with the use of steady-state potential approach (Panel A) and with dynamically changed charge (Panel B) for astronomical silicates with size $a$ = 10 nm. The sizes of computational cells are 1 au $\times$ 1 au $\times$ 1 au. The magnitude or relative statistical error is less than 5$\%$ for any point. Black curve matches the heliopause. Panel C represents the 1D distributions of ISD number density along lines $X = -11.5$, $X = - 7.5$ and $X = 1.5$ for both approaches of charge computation. Solid and dash-dotted lines denote dynamically changed and steady-state charges, correspondingly.}
    \label{figure: density_distributions_a=10nm}
\end{figure*}

In Figure \ref{figure: density_distributions_a=10nm} the analogical results are shown for particles of size $a$ = 10 nm. First, one should note that these dust particles go out from the heliosphere almost immediately after the entrance, as was seen in the corresponding panel of Figure \ref{figure: heliopause_crossing}. That is why they do not have enough time to gain a steady-state charge and, as a result, the trajectories computed with dynamically changed charge are deflected from the trajectories obtained with steady-state charge approximation. It is seen that the location and shape of density accumulation regions depend on the approach for computing the charge (panels A and B). In the case of steady-state charge approximation, there are two main explicit features in the nose part of the heliopause while, for dynamically changed charge, one can see the more elongated region of number density features. Moreover, again the trajectories computed with the use of dynamically changed charge go slightly deeper inside the heliosphere compared to the use of steady-state assumption, although the difference is relatively small ($\sim$ 1 au). The physical explanation of this effect is that the pushing electromagnetic force exerted on dust particles is proportional to the magnitude of electric charge, which is lower in the case of dynamically changed charge (Figure \ref{figure: heliopause_crossing}). In panel C the discrepancy between the two approaches is even more visible. At some points, the corresponding magnitudes of number density differ several times.

The knowledge of accurate distribution of ISD number density near the heliopause could be important in frame of planning of future Interstellar Probe mission, because the 10 nm particles are in the expected mass range of Interstellar Dust Analyzer onboard the Interstellar Probe spacecraft \citep{brandt_2023}. The same dust density features are also formed around other astrospheres, that is seen in the infrared images \citep[see, e.g.,][]{cox_2012}, and for analysing of these images the distribution of ISD number density near the astropause is used \citep{katushkina_2018}. Thus, the dynamical charge computation should be used when studying the distribution of small grains ($\sim$ 10 nm) in the vicinity of heliospheric/astrospheric boundaries.

\section{Summary and conclusions} \label{section: conclusions}

In this paper, we have presented the results of ISD charge modeling in the global heliosphere. We have considered two theoretical approaches and applied the resulting charging model to the heliospheric conditions obtained in the frame of the global 3D kinetic-MHD heliospheric model \citep{izmodenov_alexashov_2020}. The results can be summarized as follows:

\begin{enumerate}

\item The computations showed that the steady-state charge has a primarily positive value in the whole heliospheric interface. The magnitudes of steady-state charge potential inside the heliosphere are larger than in the LISM because of intensive rates of solar photons and the higher temperature of surrounding plasma. We explored the dependence of steady-state charge potential on the size of ISD particles. We showed that the steady-state potential decreases monotonically with increasing grain radius. The computational results qualitatively coincides with those presented by \cite{slavin_2012}. We also considered the effects of cosmic rays on the charging of ISD grains in the LISM and found out that it reduces the value of steady-state potential by $0.1-0.15$ V.

\item We examined the relaxation of ISD charge across the heliopause where the rapid changes of surrounding plasma parameters appear. It was shown that for biggest particles of size $a$ = 1000 nm, the relaxation of charge proceeds relatively fast ($\sim$ 0.02 years), and that is why the use of steady-state assumption in this case is justified. Medium-sized ISD particles ($a$ = 100 nm) spend approximately 0.25 years to reach an equilibrium that corresponds to the time required for ISD grain to cover the distance of 1-1.5 au. Small grains of size $a$ = 10 nm do not have enough time inside the heliosphere to approach the value of steady-state charge because they are swept away from the heliosphere by the solar wind. 

\item We performed the first simulations of ISD number density in the vicinity of the heliopause using the dynamical computation of charge, which is time-consuming. The results showed that in the case of bigger dust particles of size $a$ = 100 nm, the difference between the dynamically changed charge and steady-state charge approximation is negligible and, thus, the assumption of steady-state charge is quite valid. For smaller particles ($a$ = 10 nm), the discrepancies are more explicit in quantitatively and even qualitatively sense, and the dynamically changed charge should be used for computations in this case.

\end{enumerate}

\section*{Acknowledgments}

The work was supported by the Russian Science Foundation grant 19-12-00383. We thank I.I. Baliukin for useful discussions and for the help with the preparation of the manuscript. We also thank anonymous reviewers for careful reading of the manuscript and valuable comments.

\clearpage

\bibliographystyle{jasr-model5-names}
\biboptions{authoryear}
\bibliography{refs}

\appendix

\section{Sticking of plasma particles} \label{appendix: sticking}

\subsection{Sticking coefficient}
\label{appendix: sticking_coefficient}

We fully follow the study from \cite{weingartner_2001}. They examined the sticking of low-energy electrons on a macroscopic solid surface, performing the semi-empirical fit of the experimental results. The expressions for the sticking coefficient of electrons are as follows:

\begin{equation*}
    s_{el} =
    \begin{cases}
        0, & Z \leq Z_{min}, \\
        0.5 \left(1 - e^{-\frac{a}{l_{el}}}\right)\frac{1}{1 + e^{20 - N_c}}, & Z_{min} \leq Z \leq 0, \\
        0.5 \left( 1 - e^{-\frac{a}{l_{el}}})\right), & Z > 0,
    \end{cases}
\end{equation*}

\noindent
where $Z = Q/e$, $l_{el}$ is the electron escape length, $N_C = 468\left(\frac{a}{10^{-7} \:\: \text{cm}}\right)^3$ is the number of atoms, $Z_{min}$ is the most negative allowed charge:

\begin{equation*}
    Z_{min} = \text{int}\left(\frac{U_{\text{ait}}}{14.4 \: \text{V}}\frac{a}{\stackrel{\circ}{A}}\right) + 1,
\end{equation*}

\begin{equation*}
    -\frac{U_{\text{ait}}}{\text{V}} \approx
    \begin{cases}
        2.5 + 0.07(a/\stackrel{\circ}{A}) + 8(\stackrel{\circ}{A}/a), & \text{for astronomical silicates} \\
        3.9 + 0.12(a/\stackrel{\circ}{A}) + (\stackrel{\circ}{A}/a), & \text{for carbonaceous grains}.
    \end{cases}
\end{equation*}

Positive ions have a high probability of sticking to the surface of a dust grain because of the high values of their ionization potentials. Thus, for simplicity, we perform computations using $s_p = 1$ for all examined cases.

\subsection{Collisional cross section}
\label{appendix: collisional_cross_section}

Keeping the designations from \cite{draine_sutin_1987}, let us consider a dust grain of radius $a$, charge $Q = Ze$, and a plasma particle with charge $q_i$, kinetic energy $E_i$. If we take account of only the Coulomb interaction between them, then the interaction potential (in Gaussian units) is:

\begin{equation*}
    \Phi^0_i(Z, r) = \frac{q_i Ze}{r},
\end{equation*}

\noindent
where $r$ is the distance between them. Using the conservation of energy and angular momentum, one can obtain the following expression for the collisional cross section \citep{spitzer_1941}:

\begin{equation*}
    \sigma^0_i(Z, a, E_i) = \pi a^2\left(1 - \frac{q_i Ze}{E_ia}\right)
\end{equation*}

\noindent
Introducing dimensionless parameters

\begin{equation*}
    \varepsilon_i = \frac{E_ia}{q_i^2}, \:\:\: \nu_i = \frac{Ze}{q_i}, \:\:\: \hat{\sigma}^0_i = \frac{\sigma^0_i}{\pi a^2},
\end{equation*}

\noindent
we receive:

\begin{equation*}
    \hat{\sigma}^0_i(\varepsilon_i, \nu_i) = 1 - \frac{\nu_i}{\varepsilon_i},
\end{equation*}

\noindent
and for the case $\nu_i > \varepsilon_i$, we have $\hat{\sigma}^0_i = 0$:

\begin{equation} \label{formula: cross_section_1}
    \hat{\sigma}^0_i(\varepsilon_i, \nu_i) =
    \begin{cases}
        1 - \frac{\nu_i}{\varepsilon_i}, & \varepsilon_i \geq \nu_i, \\
        0, & \varepsilon_i < \nu_i.
    \end{cases}
\end{equation}

To be more precise, we should use the more advanced expression for the interaction potential, which was examined by \cite{draine_sutin_1987}:

\begin{equation*}
    \Phi^1_i(Z, a, r) = \frac{q_iZe}{r} - \frac{q_i^2 a^3}{2r^2(r^2 - a^2)}
\end{equation*}

If we apply mathematical transformations from Section II and Appendix B of \cite{draine_sutin_1987}, then we obtain the following expression for the dimensionless cross section:

\begin{equation} \label{formula: cross_section_2}
    \hat{\sigma}^1_i(\varepsilon_i, \nu_i) =
    \begin{cases}
        x^2 + \frac{1}{\varepsilon_i}\left(\frac{1}{2(x^2 - 1)} - \nu_i x\right), & \varepsilon_i > \theta_{\nu}, \\
        0, & \varepsilon_i \leq \theta_{\nu},
    \end{cases}
\end{equation}

\noindent
where $x > 1$ is a root of equation:

\begin{equation*}
    (2\varepsilon_i x - \nu_i)(x^2 - 1)^2 - x = 0,
\end{equation*}

\noindent
and $\theta_{\nu}$ is a dimensionless measure of the value of potential maximum, which can be expressed as:

\begin{equation*}
    \theta_{\nu} = \frac{\nu_i}{\xi_{\nu}} - \frac{1}{2\xi_{\nu}^2(\xi_{\nu}^2 - 1)},
\end{equation*}

\noindent
where $\xi_{\nu} > 1$ is the dimensionless distance at which the maximum of $\Phi^1_i(Z, a, r)$ appears and is determined as a root of equation:

\begin{equation*}
    2\xi_{\nu}^2 - 1 - \nu_i \xi_{\nu} (\xi_{\nu}^2 - 1)^2 = 0.
\end{equation*}

\begin{figure*}
    \center{\includegraphics[width=1.0\linewidth]{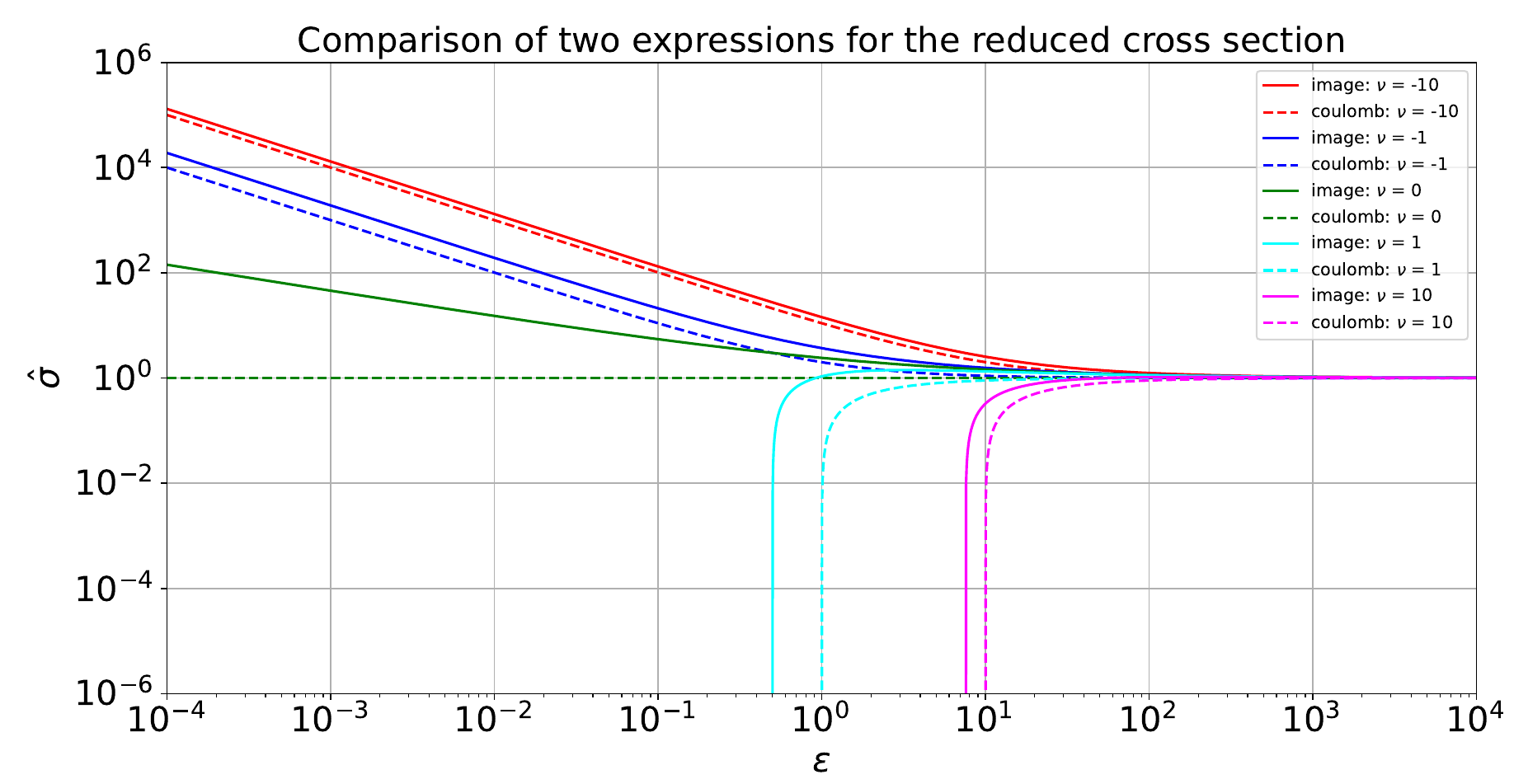}}
    \caption{Comparison of the two approaches for computing the collisional cross section. Different colors correspond to different values of $\nu$. Dashed lines are obtained using the equation (\ref{formula: cross_section_1}), and solid lines are from the equation (\ref{formula: cross_section_2}).}
    \label{figure: comparison_cross_sections}
\end{figure*}

One can find the details of derivation in \cite{draine_sutin_1987}. The difference between these two expressions for the reduced cross-section (\ref{formula: cross_section_1}), (\ref{formula: cross_section_2}) is demonstrated in Figure \ref{figure: comparison_cross_sections}. For negative values of $\nu$, the difference between corresponding curves is small, as well as for grains with high positive values of $\nu$. For grains with small positive $\nu$, the difference is noticeable. Small positive values of $\nu$ correspond to the interaction between electrons and dust grains with a slightly negative charge or between protons and grains with a slightly positive charge. Conditions favorable for small values of $\nu$ appear in the LISM. In this study, we apply a more general equation (\ref{formula: cross_section_2}) for computations.

\subsection{Dimensionality reduction of integrals}
\label{appendix: reduced_collisional_integrals}

Integral in equation (\ref{formula: sticking_general_integral}) with the Maxwellian distribution function allows for the analytical integration over $\vartheta$ and $\varphi$ variables independent on the expression for $\sigma = \sigma(v)$. After the integration, we obtain the following formula:

\begin{equation*}
    J^{st}_i = \frac{s_i q_i e n_i \pi a^2}{\hat{w}}\left(\frac{2k T_i}{\pi m_i}\right)^{1/2} \times
\end{equation*}
\begin{equation} \label{formula: reduced_sticking_rate}
    \times \int_{0}^{\infty}\hat{\sigma} \hat{v}^2 \left(\exp{\left[-(\hat{v} - \hat{w})^2\right]} -           \exp{\left[-(\hat{v} + \hat{w})^2\right]}\right) d\hat{v},
\end{equation}

\noindent
where variables and values

\begin{equation*}
    \hat{v} = \frac{v}{c_i}, \:\: \hat{w} = \frac{w}{c_i}, \:\: c_i = \left(\frac{2kT_i}{m_i}\right)^{1/2}
\end{equation*}

\noindent
were introduced for convenience. If we apply the expression (\ref{formula: cross_section_1}) for $\hat{\sigma}$, then we receive formulae analogical to the almost analytical expressions in \cite{kimura_mann_1998}. However, in this paper, we use the formula (\ref{formula: cross_section_2}) for $\hat{\sigma}$, which does not permit any promotions in the integration over $\hat{v}$ variable. This reduction of the integral dimensionality from $3$ to $1$ significantly increases the computational efficiency without accuracy losses. 

\section{Photoelectric emission}
\label{appendix: photoemission}

The photon flux $F(E)$ consists of two components: one is from the Sun, and the other is from the ISM. For the fluxes of solar photons, we use the FISM2 Daily Averages dataset\footnote{\href{https://lasp.colorado.edu/lisird/data/fism_daily_hr/}{FISM2 Daily Averages dataset}}, whose data were obtained from the SORCE/XPS (0 - 6 nm), the SDO/EVE (6 - 105 nm), the SORCE/SOLSTICE (115 - 190 nm) experiments and processed in LASP. The flux of solar photons decreases inversely proportional to the squared heliodistance, and data from the FISM2 Daily Averages dataset are obtained at $1 \: \text{au}$. We average these fluxes over time because they vary significantly along a 22-years solar cycle, but, in the future, for the needs of the time-dependent problem, these fluxes could be used as they are. For the interstellar background radiation, we follow the expressions presented in \cite{weingartner_2001}, which were based on the estimates of \cite{mezger_mathis_panagia_1982} and \cite{mathis_mezger_panagia_1983}.

\begin{figure}
    \center{\includegraphics[width=1.0\linewidth]{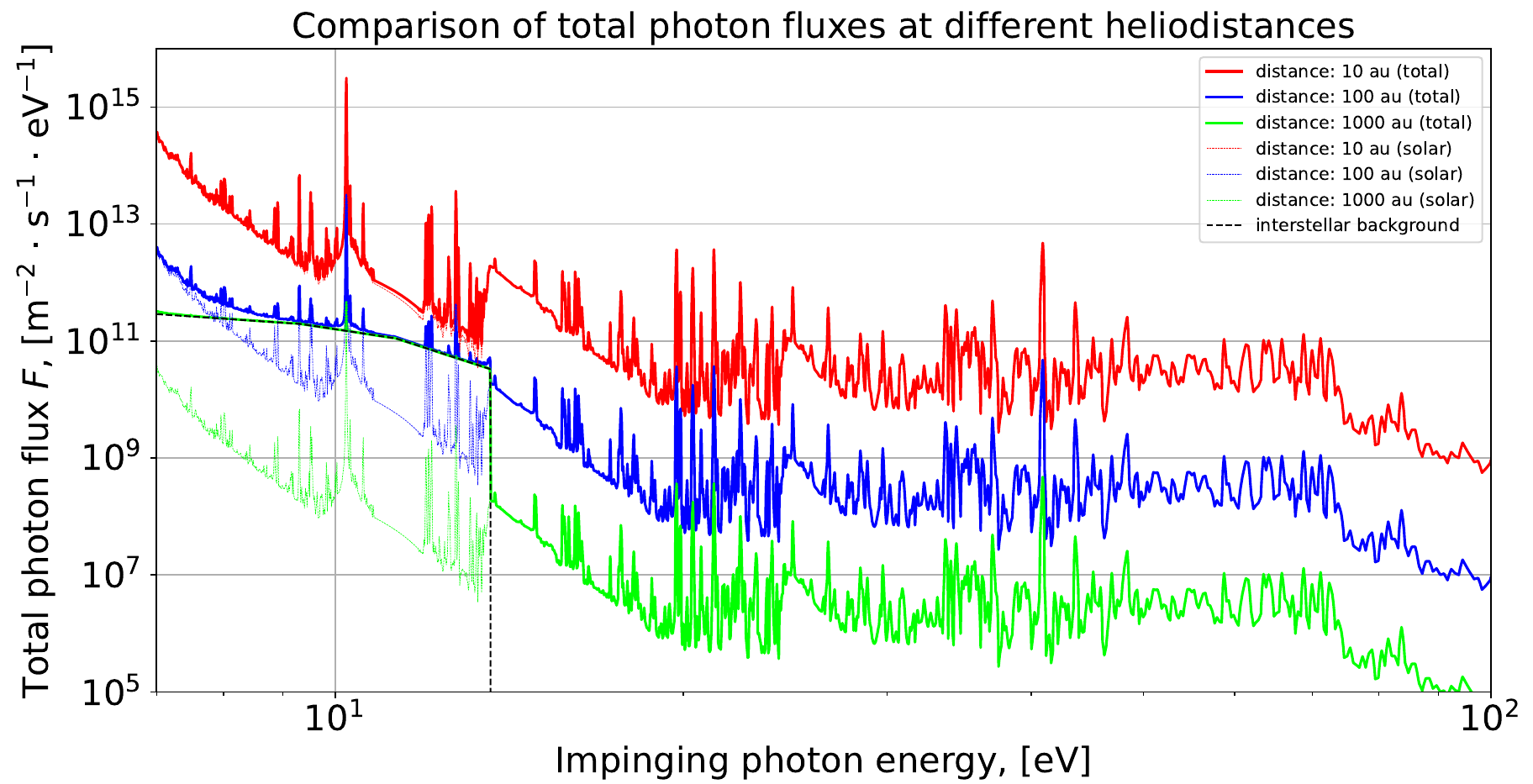}}
    \caption{The spectra of photon fluxes. Red curve is the photon flux at the heliodistance of $10$ au, blue curve -- at $100$ au, green curve -- at $1000$ au. Solid colored lines correspond to total photon fluxes, and dashed colored lines are for solar photon fluxes. Black dashed line matches the interstellar background photon flux. Close to the Sun, the dominant source of radiation that affects the charging of dust grains is the flux of solar photons, while at large heliocentric distances, it is the flux of interstellar background photons.}
    \label{figure: total_photon_flux_spectrum}
\end{figure}

Both solar and interstellar background photon fluxes should be converted into the same units. We should use [m$^{-2} \cdot$ s$^{-1} \cdot$ eV$^{-1}$] units for fluxes to be consistent with other charging currents. A dust grain is affected by the total flux of photons obtained as the sum of solar and interstellar background photons. Figure \ref{figure: total_photon_flux_spectrum} shows the spectra of photon fluxes at different heliodistances. It is clear that at large heliodistances, the flux of photons is governed by photons from the ISM, while near the Sun, the influence of interstellar background photons is almost negligible compared to the one of solar photons.

\section{Stochastic charging}
\label{appendix: stochastic_charging}

A kinetic approach is based on the assumption of statistical equilibrium \citep{gail_1975, draine_sutin_1987}. In this approach, the charge distribution function and currents are connected by the following set of balance equations:

\begin{equation} \label{formula: statistical equilibrium}
    f(Z)J_{pos}(Z) = f(Z + 1)J_{neg}(Z + 1), \:\:\: Z \in \{..., -1, 0, 1, ...\}
\end{equation}

\noindent
where $f(Z)$ is the probability that grain has a net charge $Q = Ze$, $J_{pos}$ is the sum of absolute values of currents which increase the charge: sticking of primary plasma protons, secondary electron emission, and photoemission, and $J_{neg}$ is the sum of absolute values of currents which decrease the charge: sticking of primary plasma and cosmic ray electrons.

The set of equations (\ref{formula: statistical equilibrium}) has the following solution:

\begin{equation*} \label{formula: positive_charge}
    f(Z > 0) = f(0)\prod\limits_{Z' = 1}^{Z}\left[\frac{J_{pos}(Z' - 1)}{J_{neg}(Z')}\right],
\end{equation*}
\begin{equation*} \label{formula: negative_charge}
    f(Z < 0) = f(0)\prod\limits_{Z' = Z}^{-1}\left[\frac{J_{neg}(Z' + 1)}{J_{pos}(Z')}\right],
\end{equation*}

\noindent
and constant $f(0)$ can be found by the normalization condition:

\begin{equation*}
    \sum\limits_{Z_{min}}^{Z_{max}} f(Z) = 1,
\end{equation*}

\noindent
where $Z_{max}$ and $Z_{min}$ are the most positive (the maximal charge for which an electron can be ejected) and most negative (the minimal charge for which field emission does not occur) possible charges of dust grains -- for these values, we use equations (24) and (22) from \cite{weingartner_2001}. Such a procedure allows us to obtain the charge probability distribution of dust grains.

\begin{figure*}
    \center{\includegraphics[width=1\linewidth]{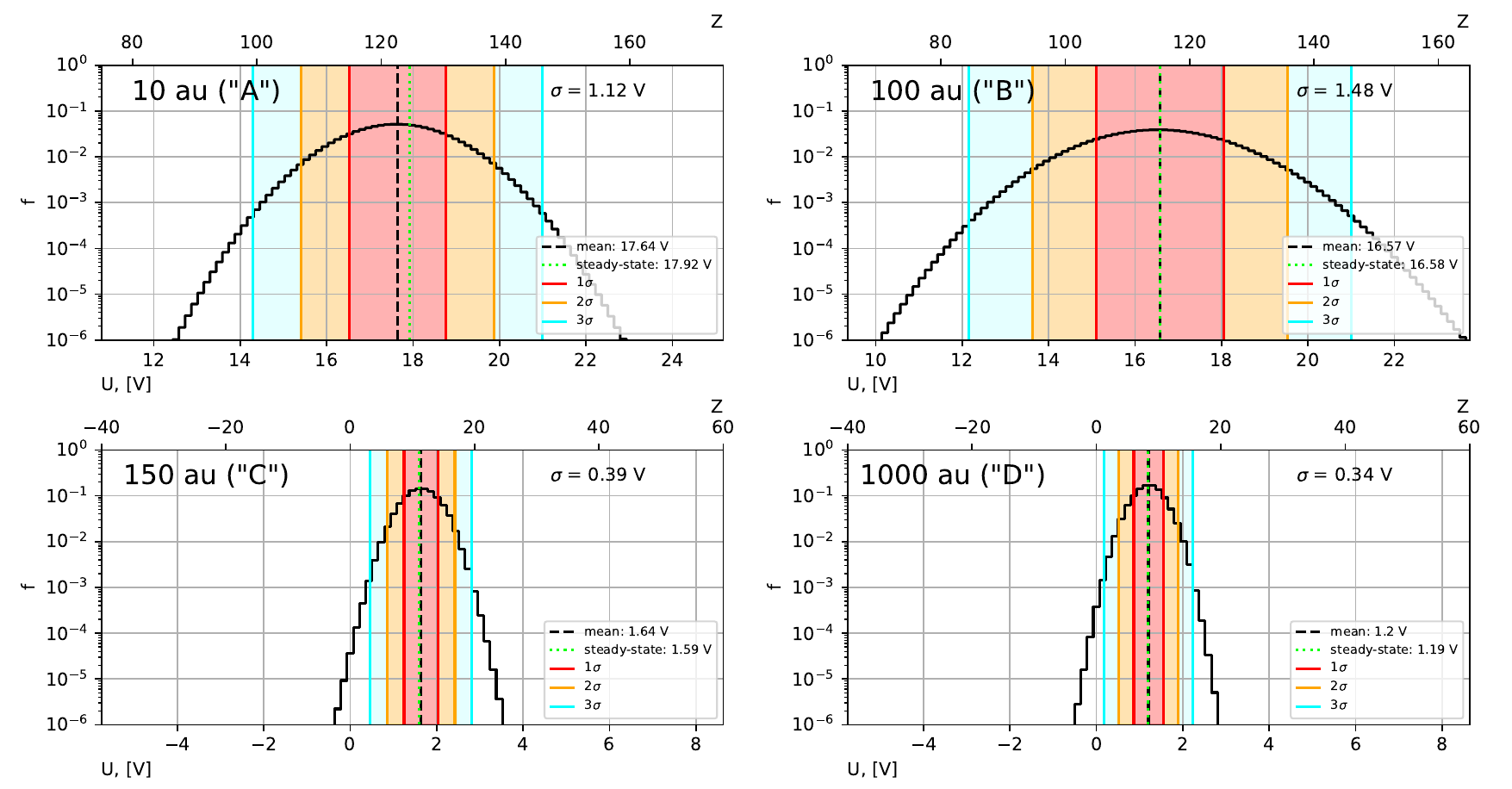}}
    \caption{The charge probability distributions of astronomical silicates obtained using the statistical equilibrium assumption. The radius of particles $a = 10$ nm. Each panel corresponds to a certain point located at the specified heliodistance along the upwind direction like in Figure \ref{figure: heliospheric_currents}. Green dashed vertical line shows the value of steady-state charge and black dashed vertical line corresponds to the mean of the distribution. Red, orange, and cyan shaded regions identify 1$\sigma$, 2$\sigma$, and 3$\sigma$ ranges within the mean, respectively, where $\sigma$ is a standard deviation.}
    \label{figure: charge_distributions}
\end{figure*}

We compute the charge probability distribution at points "A" - "D" (Section \ref{subsection: results_currents}) for astronomical silicates of radius $a = 10$ nm. The resulting distributions are presented in Figure \ref{figure: charge_distributions} with the highlighted values of steady-state charge, mean of the distribution and standard deviations.

It is seen that the difference between the steady-state charge and mean of the distribution is negligibly small for points "B", "C" and "D". For point "A", this difference is visible but still small in terms of relative differences ($\approx 1.5 \: \%$).

The ratio of standard deviation to the mean of distribution inside the heliosphere (points "A" and "B") is approximately 5-10 \%, and we do not expect that these deviations affect the trajectories and distributions of ISD grains. At the same time, outside the heliosphere (points "C" and "D"), the relative values are significantly higher (20-30 \%), and the effects on trajectories and density distributions could appear. However, it is beyond the scope of this paper and does not contradict the results and conclusions obtained in Section \ref{subsection: density_distributions}.

\section{Estimate of charge relaxation time}
\label{appendix: rough_estimate}

To provide an estimate of steady-state charge validity \citep[see, e.g.,][]{meyer-vernet_1982, kimura_mann_1998}, one can consider the relaxation time $\tau_{eq}$ defined as the time required for a dust grain to accumulate the steady-state charge starting from a neutral state:

\begin{equation} \label{formula: relaxation_time}
    \tau_{eq} = \int\limits_{0}^{Q_{eq}} \frac{dQ}{\sum\limits_k J_k(Q)}, 
\end{equation}

\noindent
where $Q_{eq}$ is the steady-state charge. Let us also introduce the characteristic time $\tau_{*}$ of ISD motion in the heliosphere \citep{kimura_mann_1998}:

\begin{equation} \label{formula: ISD_time}
    \tau_{*} = \frac{1 \: \text{au}}{26.4 \: \text{km} \cdot \text{s}^{-1}} \approx 5.68 \times 10^{6} \: \text{s}.
\end{equation}

\begin{figure}
    \center{\includegraphics[width=1\linewidth]{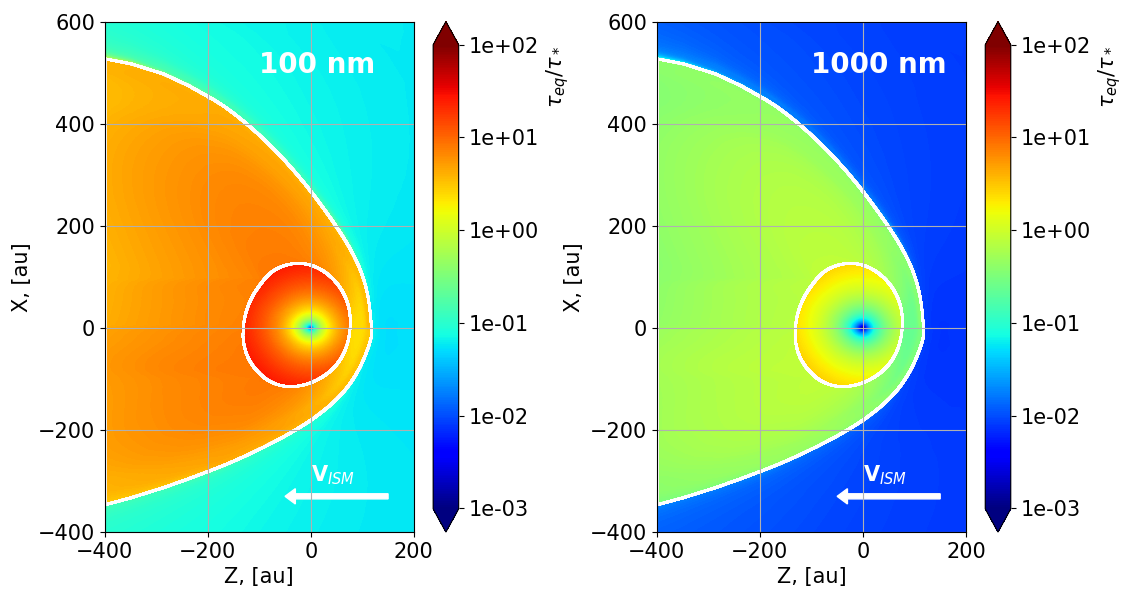}}
    \caption{The distributions of the equilibrium ratio along the plane $Y = 0$ for astronomical silicates. ({left panel}): The radius of particles is 100 nm. ({right panel}): The radius of particles is 1000 nm. White curves match the discontinuities of the heliospheric model used. Relaxation time was computed using the equation (\ref{formula: relaxation_time}).}
    \label{figure: relaxation_times}
\end{figure}

Thus, the ratio $\tau_{eq}$ to $\tau_*$ (which we call the "equilibrium ratio") equals the approximate distance (expressed in astronomical units) required for a dust grain to reach the steady-state charge starting from a neutral state. In Figure \ref{figure: relaxation_times} the distributions of equilibrium ratio are presented along the plane $Y = 0$ for astronomical silicates of two sizes: $a$ = 100 nm and $a$ = 1000 nm. It is seen from Figure \ref{figure: relaxation_times} that for big grains, the equilibrium ratio is less than (or order of) unity in the whole computational domain. Therefore, in this case, the steady-state potential should provide a good approximation when computing the trajectories. Small magnitudes of the equilibrium ratio for big grains are caused by the fact that all considered currents are proportional to $\pi a^2$.

For smaller dust particles outside the heliosphere, the equilibrium ratio is approximately $10^{-1}$. It means that one could use the steady-state potential for the modeling of ISD trajectories in the LISM. Inside the heliosphere, the equilibrium ratio is larger than the unity. Therefore, the accurate dynamical computation of charge may become necessary near the heliospheric boundaries where the value of equilibrium ratio changes rapidly.

\end{document}